\documentclass[usenatbib,traditabstract,final]{aa}
\usepackage{natbib}
\usepackage{amsmath,amsfonts,amssymb,amsxtra} % AMSTEX unterstuetzen
\usepackage{txfonts}
\usepackage{xspace}
\usepackage{graphicx}
\usepackage{natbib}
\usepackage{color}

\newcommand{\logg}{\log g }
\newcommand{\Msun}{\,$\rm{M}_\odot$\xspace}
\newcommand{\vsini}{\varv \sin i }
\newcommand{\vzams}{\varv_{\mathrm{ZAMS}} }
\newcommand{\kms}{$\rm{km/s}$\xspace}
\newcommand{\Teff}{\rm{T}_{\rm{eff}}}

\defcitealias{Brott10_popsyn}{Paper~II}

\begin{document}
\title{Rotating Massive Main-Sequence Stars I: Grids of Evolutionary Models and Isochrones}
\author{ I. Brott \inst{1,2}  \and {S. E. de Mink} \inst{3,4,9} \and {M. Cantiello} \inst{4} \and  {N. Langer} \inst{4,1} \and {A. de Koter} \inst{5,1} \and {C. J. Evans} \inst{6} \and {I. Hunter} \inst{7} \and {C. Trundle} \inst{7}  \and  {J.S.~ Vink} \inst{8}}

\institute {Astronomical Institute, Utrecht University, Princetonplein 5, 3584 CC, Utrecht, The Netherlands 
\and University of Vienna, Department of Astronomy, T\"urkenschanzstr. 17, A-1180, Vienna, Austria
\and Space Telescope Science Institute, 3700 San Martin Drive, Baltimore, MD 21218, USA
\and Argelander-Institut f\"ur Astronomie der Universit\"at Bonn, Auf dem H\"ugel 71, 53121 Bonn, Germany 
\and Astronomical Institute Anton Pannekoek, University of Amsterdam, Kruislaan 403, 1098 SJ, Amsterdam, The Netherlands
\and UK Astronomy Technology Centre, Royal Observatory Edinburgh, Blackford Hill, Edinburgh, EH9 3HJ, UK 
\and Astrophysics Research Centre, School of Mathematics \& Physics, The Queens University of Belfast, Belfast, BT7 1NN, Northern Ireland, UK
\and Armagh Observatory, College Hill, Armagh, BT61 9DG, Northern Ireland
\and Hubble Fellow 
 }

\date{Received:   Accepted: }

\abstract{We present a dense grid of evolutionary tracks and isochrones of rotating massive main-sequence stars.
We provide three grids with different initial compositions tailored to compare with early OB stars in the 
Small and Large Magellanic Clouds and in the Galaxy.  
Each grid covers masses ranging from 5 to 60\Msun and initial rotation rates between 0 and about 600\,\kms.
 
To calibrate our models we used the results of the VLT-FLAMES Survey of Massive Stars.  We determine the amount of 
convective overshooting by using the observed drop in rotation rates for stars with surface gravities 
$\log g <3.2$ to determine the width of the main sequence.  
We calibrate the efficiency of rotationally induced mixing using the nitrogen abundance determinations 
for B stars in the Large Magellanic cloud. 

We describe and provide evolutionary tracks and the evolution of the central and surface abundances.  
In particular, we discuss the occurrence of quasi-chemically homogeneous evolution,  
i.e. the severe effects of efficient mixing of the stellar interior found for the most massive fast rotators. 

We provide a detailed set of isochrones for rotating stars. Rotation as an initial parameter 
leads to a degeneracy between the age and the mass of massive main sequence stars if determined from its observed location in the 
Hertzsprung-Russell diagram. 
We show that the consideration of surface abundances can resolve this degeneracy. }

\titlerunning{Grids of Evolutionary Models and Isochrones}

\maketitle

%:----------------------------- INTRODUCTION -------------------------------------------
\section{Introduction}

Massive stars can be considered as cosmic engines. With their high luminosities,
strong stellar winds and violent deaths they drive the evolution of galaxies
throughout the history of the universe. Even before galaxies formed, massive
stars are believed to have played an essential role in re-ionizing the universe
\citep{Haiman97}, with important consequences for its subsequent evolution.
Massive stars are visible out to large distances in the nearby universe
\citep{Kudritzki08}, and as ensembles even in star-forming galaxies at high
redshift \citep[e.g.,][]{Douglas09}.  Their explosions as supernovae or
gamma-ray bursts shine through a major fraction of the universe, probing
intervening structures as well as massive star evolution at the lowest
metallicities \citep{Savaglio06, Ohkubo09}.  It is therefore of paramount
importance for many areas within astrophysics to obtain accurate models of
massive star evolution.  In this paper we present an extensive grid of
evolutionary models of rotating massive stars focusing on the main-sequence
stage, which is at the root of understanding their subsequent evolution.

Basic concepts about massive main-sequence stars are long
established and provide a good guide to their expected properties, most
importantly the mass--luminosity relation \citep{Mitalas84,
  Gonzalez05}.  However, two major issues have plagued evolutionary models of
massive stars until today: mixing and mass loss
\citep{Chiosi86}. 
 We concentrate here on the role of mixing in massive stars as,  on the main sequence, the effects of mass loss
remain limited in the considered mass and metallicity range.

Concerning thermally driven mixing processes,
the occurrence of semiconvection and the related question of the appropriate choice of the
criterion for convection do not alter the main sequence evolution of massive stars
significantly \citep{Langer85}. However, the efficiency of convective overshooting
in massive main-sequence stars is still not well known. In particular, the cool edge of the
main-sequence band is not well determined, as stars in observational samples are found in
the gap predicted between main-sequence stars and blue core-helium burners
\citep[e.g.][]{Fitzpatrick90,Evans06_lmc_smc}.
Here we follow a promising approach to determine the cool edge of the
main sequence band, and thus the overshooting efficiency, from the rotational properties
of B-type stars \cite[see][]{Hunter08_vrot,Vink10}.

Rotationally-induced mixing processes
have been invoked in the past decade to explain the surface enrichment
of some massive main-sequence stars with the products from hydrogen burning,
in particular nitrogen \citep{Heger00, Meynet00, Maeder00}.
Rotationally-induced mixing was further used to explain the
ratio between O-stars and various types of Wolf-Rayet stars at different
metallicities \citep{MeynetMaeder05_wolfrayetpopulations},
the variety of core collapse supernovae
\citep{Georgy09} and the evolution towards long gamma-ray bursts \citep{Yoon05}.
However, in view of the fact that all the mentioned observed phenomena have 
alternative explanations, a direct test of rotational mixing in massive stars 
appears of paramount importance.

In view of the recent results from the FLAMES Survey of 
Massive Stars \citep{Evans05_gal}, which provided spectroscopic data for
large samples of massive main sequence stars, we pursued the strategy of  
computing dense model grids of rotating massive main sequence stars,
and predict in detail and comprehensively their observable properties. 
This evolutionary model data~\footnote{Our model data is made available through VizieR
at \url{http://cdsarc.u-strasbg.fr/cgi-bin/...}}, 
which was calibrated using results from the FLAMES Survey, is tailored to provide the first 
direct and quantitative test of rotational mixing in rapidly rotating stars\citep[][from here on Paper~II]{Hunter08_letter, Hunter09_nitrogen,Brott10_popsyn}.

Our new model grids deliver predictions for many observables, with a wide range of potential applications. 
Besides helium and the CNO elements, we show how the light elements lithium, beryllium and boron
are affected by rotational mixing, as well as the elements fluorine and sodium.
We also provide isochrones which
predict complex structures near the turn-off point of young star clusters.
In Sec.~\ref{sec:bec}, we describe the set-up and calibration of our massive star models,
Sec.~\ref{sec:evolgrid} and~\ref{sec:abund} provide the evolutionary tracks in the HR diagram and
the corresponding isochrones, and the surface abundances of all considered elements.
We end with a brief summary in Sec.~\ref{sec:summary}.

%:-------- Stellar Evolution Models --------
\section{Stellar evolution code}
\label{sec:evolution_models} 
\label{sec:bec}

We use a one-dimensional hydrodynamic stellar evolution code that takes into account the physics of rotation, 
magnetic fields and mass-loss. This code has been described extensively by \citet{Heger00}.  
Recent improvements are presented in \citet{Petrovic05_GRB} and \citet{Yoon06}. 

To allow for deviations from spherical symmetry due to rotation in this one-dimensional code, we consider the stellar properties and structure equations on mass shells that correspond to isobars. According to  e.g. \citet{Zahn92} turbulence efficiently erases gradients along isobaric surfaces and enforces shellular rotation \citep{MeynetMaeder97} allowing us to use the one-dimensional approximation.   The effect of the centrifugal acceleration on the stellar structure equations is considered according to \citet{KippenhahnThomas70} and \citet{EndalSofia76} as described in \citet{Heger00}. 

\subsection{Transport of chemicals and angular momentum}

All mixing processes are treated as diffusive processes.  Convection is modeled using the Ledoux criterion, adopting a mixing-length parameter of $\alpha_{\rm{MLT}}=1.5$ \citep{BohmVitense58,Langer91}. Semi-convection is treated as in \citet{Langer+83} adopting an efficiency parameter $\alpha_{\rm{SEM}}=1$ \citep{Langer91}. We take into account convective core-overshooting using an overshooting parameter of $0.335$ pressure scale heights  i.e. the radius of the convective core is equal to the radius given by the Ledoux criterium plus an extension equal to 0.335 Hp, where Hp is the pressure scale height evaluated at the formal boundary of the convective core (see Sec.~\ref{sec:overshooting} for the calibration). 
Furthermore, we consider various instabilities induced by rotation that result in mixing: Eddington-Sweet circulation, dynamical and secular shear instability, and the Goldreich-Schubert-Fricke instability \citep{Heger00}.  

Transport of angular momentum is also treated as a diffusive process following \citet{EndalSofia78} and \citet{Pinsonneault89} as described in \citet{Heger00}. The turbulent viscosity is determined as the sum of the diffusion coefficients for convection, semi-convection, and those resulting from rotationally induced instabilities.  In addition, we take into account the transport of angular momentum by magnetic fields due to the Spruit-Tayler dynamo \citep{Spruit02}, implemented as described in \citet{Petrovic05_GRB}.   We do not consider possible transport of chemical elements as a result of the Spruit-Tayler dynamo because its validity is still controversial \citep{Spruit06}. 
We note that, while the dynamo process itself was confirmed through numerical calculations by
 \citet{Braithwaite06}, it is still unclear under which conditions it can operate 
and whether it is active in massive stars \citep{Zahn07}. Heuristically, angular momentum transport through magnetic fields produced by the
Spruit-Tayler dynamo, as the so far only mechanism, appears to reproduce the rotation rates of white dwarfs
and neutron stars quite well \citep{Heger05, Suijs08}.  

In agreement with \citet{MaederMeynet05} we find that magnetic fields keep the star near rigid rotation throughout  its main sequence evolution, suppressing the mixing induced by shear between neighboring layers. The dominant rotationally induced mixing process in our models is the Eddington-Sweet circulation, a large scale meridional current that is caused by thermal imbalance in rotating stars between pole and equator \citep[e.g.][]{Tassoul78}.  

Some of the diffusion coefficients describing rotational mixing are based on order of magnitude estimates of the relevant time and length scales \citep{Heger00}.  To consider the uncertainties efficiency factors are introduced, which need to be calibrated against observational data.  The contribution of the rotationally induced instabilities to the total diffusion coefficient is reduced by a factor $f_c =0.0228$ (see Sec.~\ref{sec:rotmix} for the calibration) while their full value enters in the expression for the turbulent viscosity \citep[see][for details]{Heger00}.  The inhibiting effect of chemical gradients on the efficiency of rotational mixing processes is regulated by the parameter $f_\mu$, see \citet{Heger00}. We adopt $f_\mu=0.1$ after \citet{Yoon06} who calibrated this parameter to match observed surface helium abundances.

\subsection{Mass loss}
We updated the treatment of mass loss by stellar winds by implementing the prescription of \citet{Vink00, Vink01},
based on the method by \citet{deKoter97}, for winds from early O- and B-type stars. This mass loss recipe predicts a fast increase of the mass-loss rate as one moves to lower temperatures near $22\,000\,$K.  This increase is related to the recombination of Fe\,{\sc iv} to Fe\,{\sc iii} at the sonic point and is commonly referred to as the bi-stability jump. These rates are derived for $12.5~\mathrm{kK} \la T_\mathrm{eff} \la 50~\mathrm{kK}$. We note that the formulae on the hot and cool side of the
jump have {\it not} been derived for the intermediate temperature range between 22.5 and 27.5 kK. In this range, we thus perform a linear interpolation. 
In order to accommodate for a strong mass-loss increase when approaching the HD limit,
we switch to the empirical mass loss rate of \citet{Nieuwenhuijzen90}, when the Vink et al rate becomes smaller than that from \citet{Nieuwenhuijzen90} at any temperature lower than the critical temperature for the bi-stability jump.  This ensures a smooth transition between the two mass loss prescriptions.  
This strategy also naturally accounts for the increased mass loss at
the {\it second} bi-stability jump at $\sim$12.5 KK. 

To account for the effects of surface enrichment on the stellar winds, we follow the approach of \citet{Yoon06}. The mass-loss rate of \citet{Vink00, Vink01} is employed for stars with a surface helium mass fraction, $Y_\mathrm{s}$, below 0.4.  We interpolate between the mass-loss rate of \citet{Vink00,Vink01} and the Wolf-Rayet mass-loss rate of \citet{Hamann95} reduced by a factor of 10 \citep{Yoon06} for $0.4 \le Y_\mathrm{s} \le 0.7$.  When  the helium surface abundance exceeds $Y_\mathrm{s} > 0.7$, we adopt the Wolf-Rayet mass-loss rate. 

Rotational mass loss enhancement is included according to \citet{Yoon05}.

%:----------------------  CHEMICAL COMPOSITION --------------------------
\subsection{The initial chemical composition}
\label{sec:chem}

We adopt three different initial compositions that are suitable for comparison with OB stars in the Small and Large Magellanic Cloud (SMC, LMC) and a mixture that is tailored to the Galactic sample of the FLAMES survey of massive stars \citep{Evans05_gal}, to which we will refer in this paper as the Galactic (GAL) mixture for brevity. 
In contrast with several previous studies we do not simply adopt Solar-scaled abundance ratios. Even though such mixtures may be sufficient to study the overall effects of metallicity, they are not accurate enough for direct comparison with observed surface abundances.  For example, \citet{Kurt98} find that the carbon to nitrogen ratio in HII regions in the LMC and SMC is considerably larger than the solar ratio.   In the stellar interior, carbon is converted into nitrogen, which can reach the surface due to mixing and/or mass loss.  \citet{Brott08} found that the surface nitrogen abundance can be enhanced by a factor of 11 with a realistic initial SMC mixture. A Solar-scaled mixture with the same iron abundance would lead to enhancement of only a factor of 4, for the same initial parameters. This demonstrates the need for tailored initial chemical mixtures. 

 \begin{table}[htbp]
\caption{Initial abundances for C, N, O, Mg, Si, Fe adopted in our chemical compositions for the Magellanic Clouds and the Galaxy (see Sec.~\ref{sec:chem}).}
   \centering
 	\begin{tabular}{l | l l l | l l l }
 	\hline \hline
       	& {C} & {N} & {O} & {Mg} & {Si} & {Fe}\\	
	\hline
  	LMC & 7.75 & 6.90 & 8.35 & 7.05 & 7.20 & 7.05 \\  
  	SMC & 7.37 & 6.50 & 7.98 & 6.72 & 6.80 & 6.78\\   
  	GAL & 8.13 & 7.64 & 8.55 & 7.32 & 7.41 & 7.40 \\  
  	\hline
 	\end{tabular}
	\tablefoot{All other elements are solar abundances \citep{Asplund05} scaled down by 0.4\,dex for the LMC and by 0.7\,dex for the SMC.}
   \label{tab:compositions} 
\end{table}

For the initial abundances of C, N and O we use data from HII regions by \citet{Kurt98}. Surface abundances for these elements in B-type stars may no longer reflect the initial composition due to nuclear processing and mixing.  
For Mg and Si we use measurements of B-type stars as these stars are not expected to significantly alter the abundances of these elements during their main-sequence evolution. The B-type stars in different fields and clusters in the LMC  show only small variations in Mg and Si of less than 0.02 dex \citep{Trundle07, Hunter07_chem_Bstars, Hunter09_nitrogen}. Also for the SMC these authors find no evidence for any systematic differences. The Fe abundance for the SMC is taken from a study of A-supergiants by \citet{Venn99}. For all the other elements we adopt solar abundances by \citet{Asplund05} scaled down by 0.4 dex for the LMC mixture and 0.7 dex for the SMC mixture.  The initial abundances are summarized in Tab.~\ref{tab:compositions}.  These mixtures have metallicities of $Z = 0.0047$ for the LMC and $Z=0.0021$ for the SMC, where $Z$ is defined as the mass fraction of all elements heavier than helium (see Table~\ref{tab:Z}).

Choosing a chemical composition suitable for comparison with stars in the Milky Way is not straight forward due to the chemical gradients within the disk of our galaxy. We adopt an initial Mg and Si abundance based on the analysis of the Galactic cluster NGC6611 by \citet{Hunter07_chem_Bstars}. 
For C we adopted the NLTE corrected value from \citet{Hunter07_chem_Bstars}.  N and O abundances are based on the average of HII-regions, see \citet{Hunter08_vrot,Hunter09_nitrogen} and references therein. The Fe abundance is taken from measurements of A-supergiants by \citet{Venn95}. For all other elements we have adopted solar abundances by \citet{Asplund05}. The resulting metallicity of our Galactic mixture is $\rm{Z}=0.0088$,
which is lower than the solar metallicity of $\rm{Z}=0.012$ found by \citet{Asplund05} and the solar neighborhood metallicity
of $\rm{Z}=0.014$ measured by \citet{Przybilla08}.
We point out we assumed the stellar wind mass loss rates as well as the opacities in our models to depend on the iron 
abundance, for which the spread in measurements is much smaller than for the total metallicity. 
Our wind mass loss rate scales with $(\rm{Fe}_{\rm{Surf}}/\rm{Fe}_{\odot})^{0.85}$, 
where we use the iron abundance from \citet{Grevesse96} as solar reference, for consistency reasons with older models.

For helium we assume that the mass fraction scales linearly with Z between the primordial helium 
mass fraction of $\rm{Y}=0.2477$ \citep{Peimbert07} at $\rm{Z}=0$ and $\rm{Y}=0.28$ at the 
solar value \citep{Grevesse96}.  We adopt the OPAL opacity tables \citep{opal} using 
$(\rm{Fe}_{Surf}/\rm{Fe}_{\odot})\times \rm{Z}_{\odot}$ 
to interpolate between tables of different metallicities. The solar reference values are again taken from \citet{Grevesse96}.

\begin{table}[htb]
\caption{Resulting hydrogen (X), helium (Y) and metal (Z) mass fractions for the chemical mixtures used in our models. }
\centering
\begin{tabular}{l | l l l}
\hline \hline
& X & Y & Z\\
\hline
LMC & 0.7391 & 0.2562 & 0.0047 \\
SMC & 0.7464 & 0.2515 & 0.0021\\
GAL & 0.7274 & 0.2638 & 0.0088\\
\hline
\end{tabular}
\label{tab:Z}
\end {table}

%:--------------------- OVERSHOOTING ---------------------------------------
 \subsection{Calibration of overshooting}
 
 \label{sec:overshooting}
  
 \begin{figure}[htbp]
\centering
\includegraphics[angle=-90,width=0.5\textwidth]{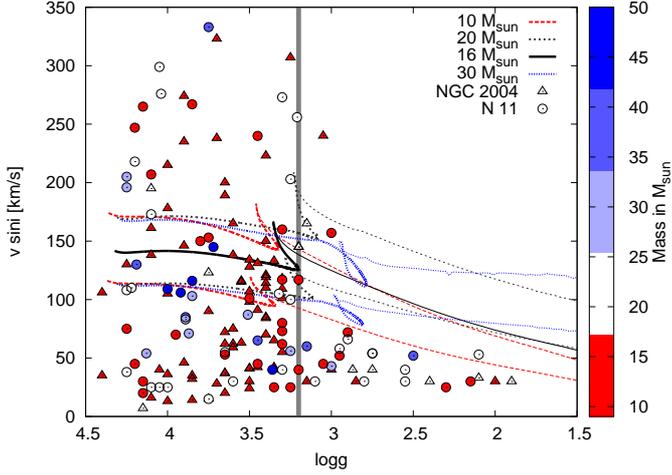}
\caption{Projected rotational velocity versus surface gravity for stars in the FLAMES survey near LMC clusters NGC 2004 (triangles) and N11 (circles). Color coding indicates the stellar mass.  We use  the sudden transition at $\logg=3.2$ (vertical line) to calibrate the amount of overshooting of our 16\Msun model (black line), see Sec.~\ref{sec:overshooting}.  For comparison we plot several other models with different initial masses and rotation rates (dotted lines). 
}
\label{fig:logg-vsini}
\end{figure}

\begin{figure}[htbp]
\begin{center}
\includegraphics[angle=-90,width=0.5\textwidth]{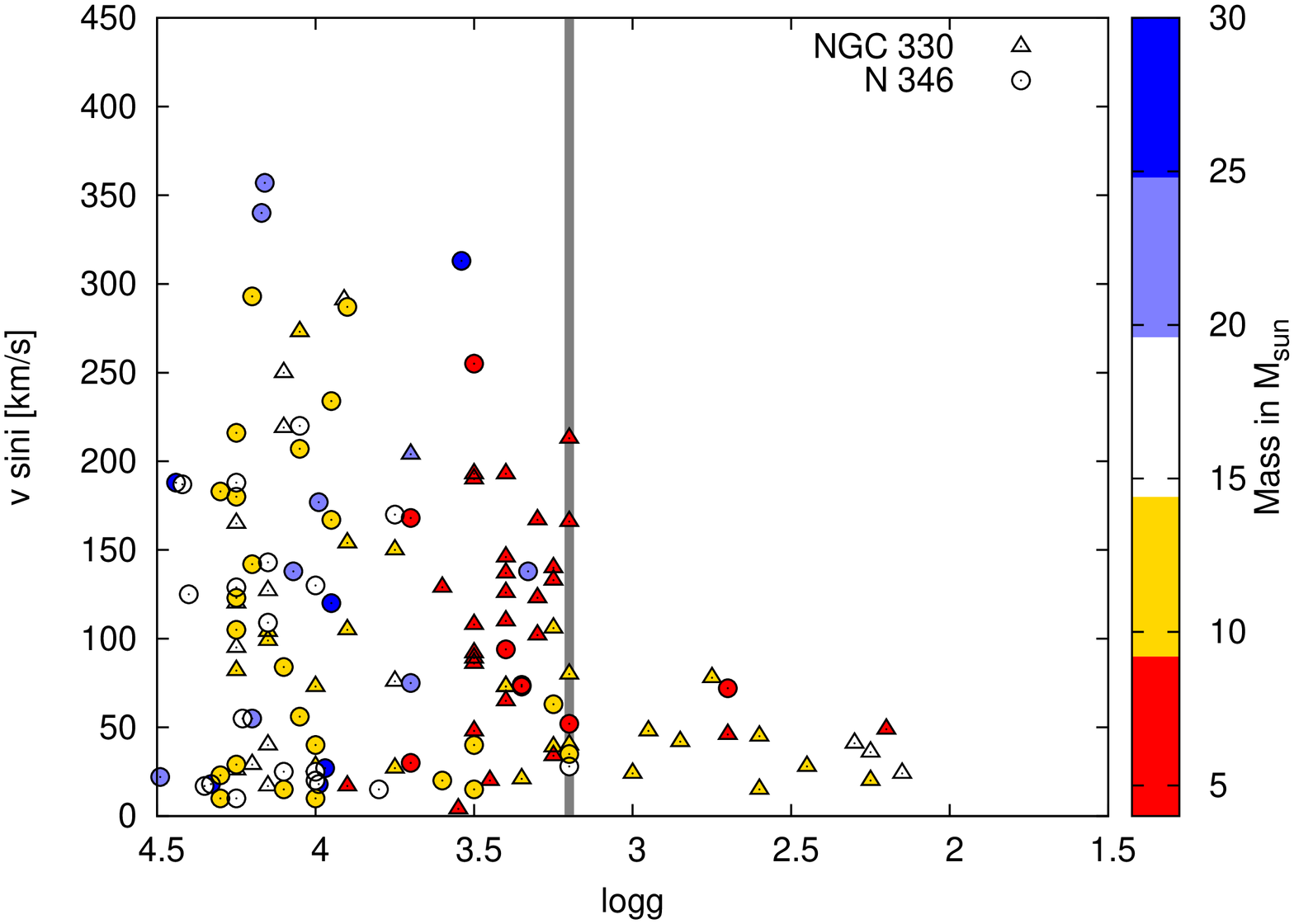}
\includegraphics[angle=-90,width=0.5\textwidth]{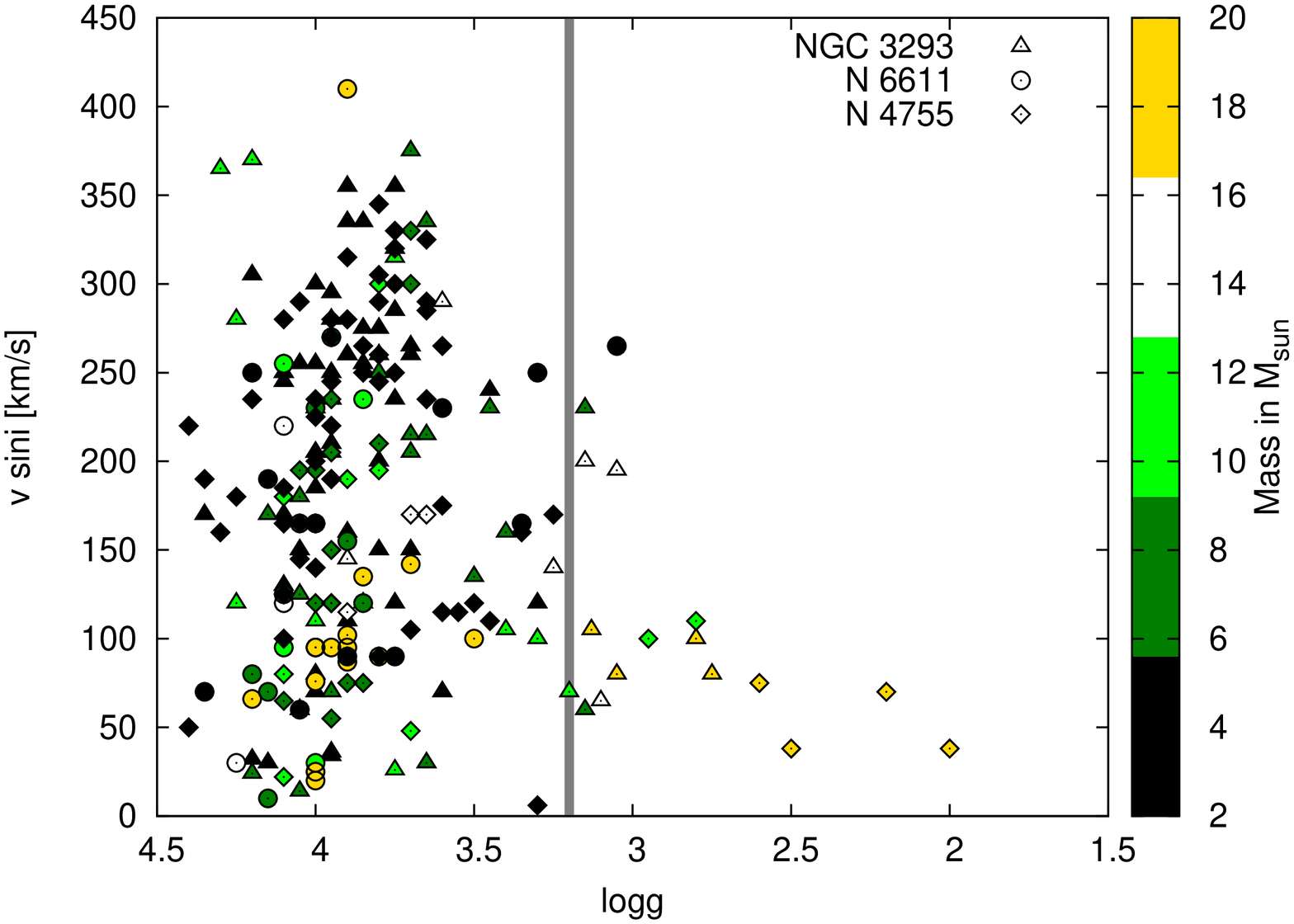}
\caption{Similar to Fig.~\ref{fig:logg-vsini}, but now for the SMC (top) and Galactic (bottom) samples of the FLAMES survey. The typical error in $\logg$ is about 0.15 dex. The drop in rotation rate, which may be interpreted as the end of the main sequence, occurs around $\logg = 3.2$ (vertical line) independently of metallicity. }
\label{fig:logg-vsini-mw-smc}
\end{center}
\end{figure}

Mixing beyond the convective stellar core can occur when convective cells penetrate into the radiative region, when they ``overshoot'' the boundary due to their non-zero velocity. This effect is accounted for through a parameter $\alpha$ which measures the extension of the affected region in units of the local pressure scale height. In general, overshooting leads to larger stellar cores resulting in higher luminosities and larger stellar radii, especially towards the end of the main sequence evolution.  

Various attempts have been undertaken to constrain the value of this parameter.  Using eclipsing binary stars, \citet{Schroeder97} derived values between 0.25 and 0.32 for stars in the mass range of 2.5 to 7\Msun. \citet{Ribas00} and \citet{claret07} found $\alpha$ in the range of 0.1 to 0.6, with a systematic increase of the amount of overshooting with the stellar mass.   \citet{Briquet07} used asteroseismological measurements and obtained $\alpha = 0.44\pm 0.07$ for the $\beta$-Cephei $\theta$ Ophiuchi.  
Another way to constrain the overshooting parameter is to fit the width of the 
main-sequence band for clusters with turn-off masses below approximately 15\Msun\citep{Mermilliod86}. 
Using non-rotating evolutionary models, \citet{Mermilliod86} and \citet{MaederMeynet87} 
found an overshooting parameter between 0.25 and 0.3. This is supported by findings of \citet{Napiwotzki91}. 
In general, however, this method does not work well for masses above 5...10\Msun, since
the red edge of the main sequence band remains mostly unidentifiable \citep{Vink10}.

In this work we re-calibrate the overshooting parameter for our models using data of the FLAMES survey of massive stars.   We use the fact that the properties of stars  near the end of their main-sequence evolution depend on their core size.  For larger overshooting parameters, the end of the main sequence shifts to lower effective temperatures and lower surface gravities. After the end of their main sequence evolution, the rapid stellar expansion results in a strong spin down of the envelope. This phase of evolution occurs on the Kelvin-Helmholtz timescale, which is very short compared to the nuclear timescale. Observing stars during this phase is unlikely. 

In Fig.~\ref{fig:logg-vsini} we plot the projected rotational velocity against the surface gravity for stars in the LMC sample of the FLAMES survey \citep[adapted from ][]{Hunter08_vrot}.  Stars with surface gravities of $\log g > 3.2$ show a wide range of projected rotational velocities, $\vsini$, in clear contrast with the stars with lower surface gravities which have projected velocities of about 50\,\kms or lower.  We interpret this transition as the division between main-sequence stars that are born with a wide range of rotational velocities and a different, more evolved, population that has experienced significant spin down. 
The surface gravity at which this transition occurs seems to be independent of mass, at least within the range available in this sample. To show this we color-coded the masses of the observed stars in Fig.~\ref{fig:logg-vsini}.

In an alternate scenario discussed by\citet{Vink10}, the population of slow rotators might be main-sequence stars, 
that have spun down due to increased mass loss (bi-stability breaking). However, this would require a huge
overshooting, in order to extend the lifetime of main sequence stars at temperatures below the bi-stability temperature
of about 22\,kK. With the current overshooting calibration, bi-stability breaking plays a role only well above 30\Msun. 
The nature of the evolved slow rotators in Fig.~\ref{fig:logg-vsini} remains puzzling. 
The fact that all of them have a strong nitrogen surface enhancement might propose that 
they are post-red supergiants which acquired the nitrogen enhancement through the first dredge-up
\citep{Vink10}, but they might as well be products of binary interaction or of so far unidentified physical processes.

For the overshooting calibration we use a stellar model with a mass and rotation rate that are representative for the entire sample. The typical mass is about 16\Msun \citepalias[based on evolutionary masses; see][]{Brott10_popsyn}  and the mean projected velocity of the stars in the sample with $\logg \geq 3.2 {\rm dex}$ is $\langle \vsini \rangle =110\,\kms$. This corresponds to an average rotational velocity of $\langle v \rangle=142\,\kms$ assuming a random orientation of the spin axes, i.e. $\langle \sin i \rangle=\pi/4$.  Throughout the main-sequence evolution the equatorial velocity is expected to be nearly constant due to the reduced effect of stellar winds in the low metallicity environment of the LMC and due to coupling of the expanding envelope to the contracting core. Therefore, we use the typical velocity as an estimate of the average initial rotational velocity. 

The full black line in Fig.~\ref{fig:logg-vsini} shows a model with the typical mass and rotation rate. The overshooting parameter has been adjusted to  $\alpha=0.335$, such that the end of the main sequence coincides with the drop in rotational velocities at $\logg=3.2$.   
Fig.~\ref{fig:logg-vsini} demonstrates that the adopted amount of overshooting, derived for a 16\Msun star, 
appears to be valid for the mass range 10 to 20 \Msun. However, it is unconstrained outside this range, as
for example the 30 \Msun tracks in Fig.~\ref{fig:logg-vsini} have no observed counterparts 
near the TAMS. We therefore use an overshooting parameter of $\alpha=0.335$ for our entire grid. 
Given the uncertainty in the surface gravities ($\sim$0.15 dex), we estimate the uncertainty 
of the overshooting parameter in the mass range 10 to 20 \Msun to be $~0.1$.

A priori we do not expect the overshooting parameter to depend on metallicity. To check this we show the projected rotational velocities against surface gravity for the SMC and Galactic sample of the FLAMES survey in Fig.~\ref{fig:logg-vsini-mw-smc}.  The drop in rotation rate occurs at the same surface gravity within the measurement errors.   
The Galactic sample shows a  second feature around $\logg=3.5$ caused mainly by stars below 5\Msun. \citet{Dufton06_vrot} already described this stars as over-luminous for their age and a general disappointing agreement of the clusters HR-diagram with isochrones.
When we vary the initial composition from SMC to the Galactic mixture, we find that the end of the main sequence for our 16\Msun model shifts from $\logg \sim 3.3 $ to 3.1, adopting the same value for the overshooting parameter. This shift is also within the measurement uncertainties. Therefore, adopting one fixed value for the overshooting parameter in our entire model grid is a reasonable assumption.

%: ------------------ MIXING EFFICIENCY ----------------------------
 \subsection{Calibration of the rotational mixing efficiency}
 \label{sec:rotmix}
 \begin{figure}[htbp]
\begin{center}
\includegraphics[angle=-90, width=0.5\textwidth, bb=70 55 546 760, clip]{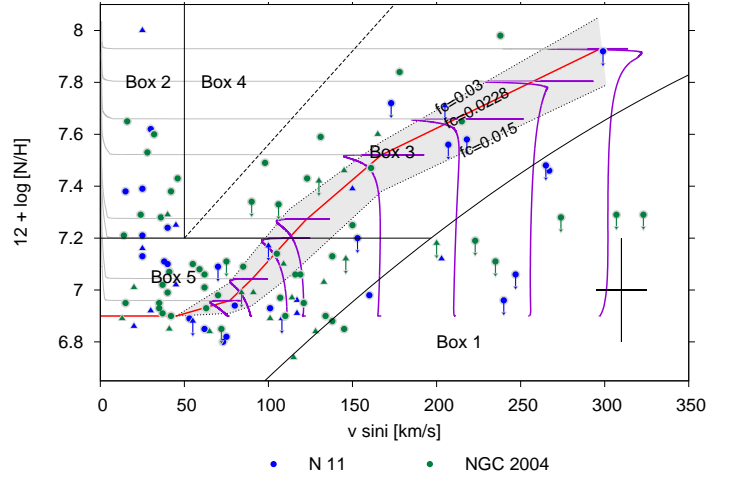}
\caption{Nitrogen surface abundance versus projected rotational velocity for stars in the LMC sample of the FLAMES survey \citep{Hunter09_nitrogen} in the N11 (blue) and the NGC\,2004 (green) fields. Stars with evidence for a binary companion from radial velocity variations are plotted as triangles.  Black solid lines indicate regions that are helpful as a reference in later discussion\citepalias[see also][]{Brott10_popsyn}.
Evolutionary tracks of 13\Msun models are plotted for various rotational velocities, shifted by a factor $\pi/4$ to account for random inclination angles.
The red line marks the nitrogen abundance reached at the end of the main sequence  for 13\Msun models adopting a mixing efficiency of $f_{c}=0.0228$ (our calibration). Thin dotted lines show the location of the line, if different values for $f_{c}$ are adopted, e.g. $f_{c}=0.015$ (lower line) and $f_{c}=0.03$ (upper line).  }
\label{fig:hunterplot}
\end{center}
\end{figure}

As mentioned in the beginning of Sec.~\ref{sec:bec} we treat all mixing processes as diffusive processes, where the diffusion coefficient is taken as the sum of the individual diffusion coefficients for the various mixing processes.    The contributions of all rotationally induced mixing processes are reduced by a factor, $f_c$, before they are added to the total diffusion coefficient for transport of chemicals.
\citet{Pinsonneault89} was the first to introduce this parameter. In order to explain the solar lithium abundance they needed to reduce the efficiency of rotationally induced mixing processes by a factor $f_c=0.046$. Based on theoretical considerations \citet{ChaboyerZahn92} proposed $ f_c = 0.033$, which was adopted in the models of  \citet{Heger00}.  
We note that we cannot compare our efficiency parameter directly with these values due to differences 
in the implementation and the considered specific set of mixing and angular momentum transport processes. 

We calibrate the mixing efficiency for our models using the B-stars in the LMC sample of the FLAMES survey. We aim to reproduce the main trend of the observed nitrogen surface abundances with the projected rotational velocity for the nitrogen enriched fast rotators. In Fig.~\ref{fig:hunterplot} we present the data, indicating different regions or boxes which have been numbered for reference (\citealp[][]{Hunter08_letter} and \citetalias{Brott10_popsyn}). The majority of the sample, about 53\% of the stars, is not significantly enriched in nitrogen, given the typical error of 0.2 dex for the nitrogen abundance measurements. This group is indicated as Box~5 in Fig.~\ref{fig:hunterplot}.  Fast rotators with significantly enhanced surface abundances ($> 7.2$ dex; see Box~3) constitute 18\% of the sample. 
 A detailed discussion of the stars in Box~1 and 2, which deviate from the main trend, and of observational biases and other factors that might influence this numbers are provided in \citetalias{Brott10_popsyn} and \citet{Hunter08_letter}. 

For the calibration we use stellar models with a mass that is typical for the stars for which we have nitrogen surface abundance measurements,  i.e. 13\Msun.  We adjust the mixing efficiency such that the nitrogen surface abundances reached at the end of the main-sequence evolution follow the trend of the stars in Box~3 while we make sure that the slow rotators do not enrich more than 0.2 dex to match the stars in Box~5.  We find that this is obtained for a mixing efficiency of $f_c$=0.0228.  In Fig.~\ref{fig:hunterplot} we plot evolutionary tracks for several 13\Msun models with different initial rotation rates.  Adopting this mixing efficiency, a model with the typical mass and average rotational velocity ($\vzams =142\,\kms$) reaches a surface nitrogen enhancement of 7.2 dex at the end of its main sequence.  
To illustrate the sensitivity of the nitrogen surface abundances to the adopted mixing efficiency, we depict the surface abundances reached at the end of the main sequence for $f_c = 0.015$ and $f_c=0.03$ (dotted lines and grey band).  
As we have no reason to believe that the mixing efficiency $f_c$  depends on metallicity, we adopt the same value for the SMC and the Galactic grid.

\begin{figure*}[!htp]
\begin{center}
\includegraphics[bb=3 25 480 732, clip, angle=0,width=.31\textwidth]{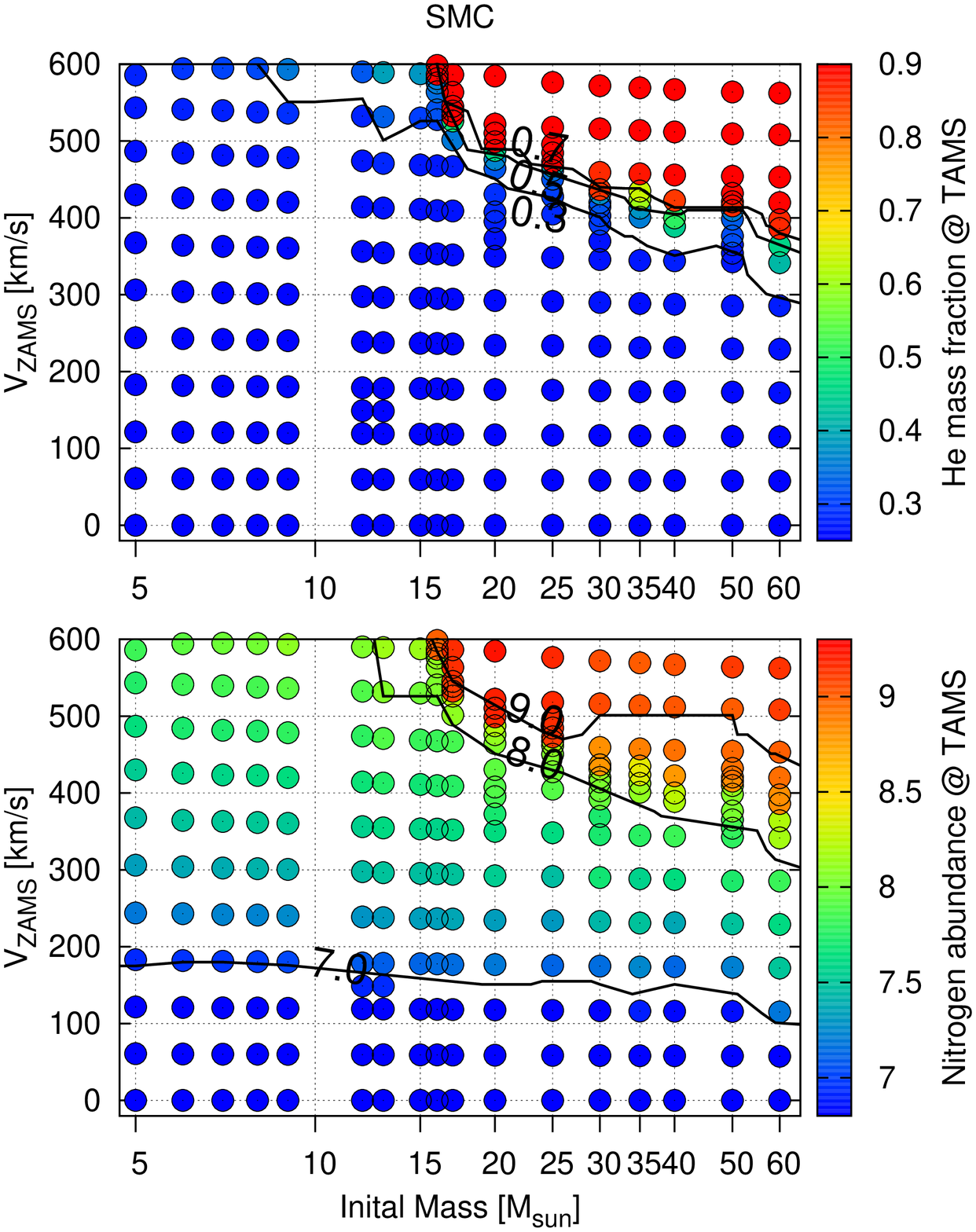}%
\includegraphics[bb=3 25 480 732, clip, angle=0,width=.31\textwidth]{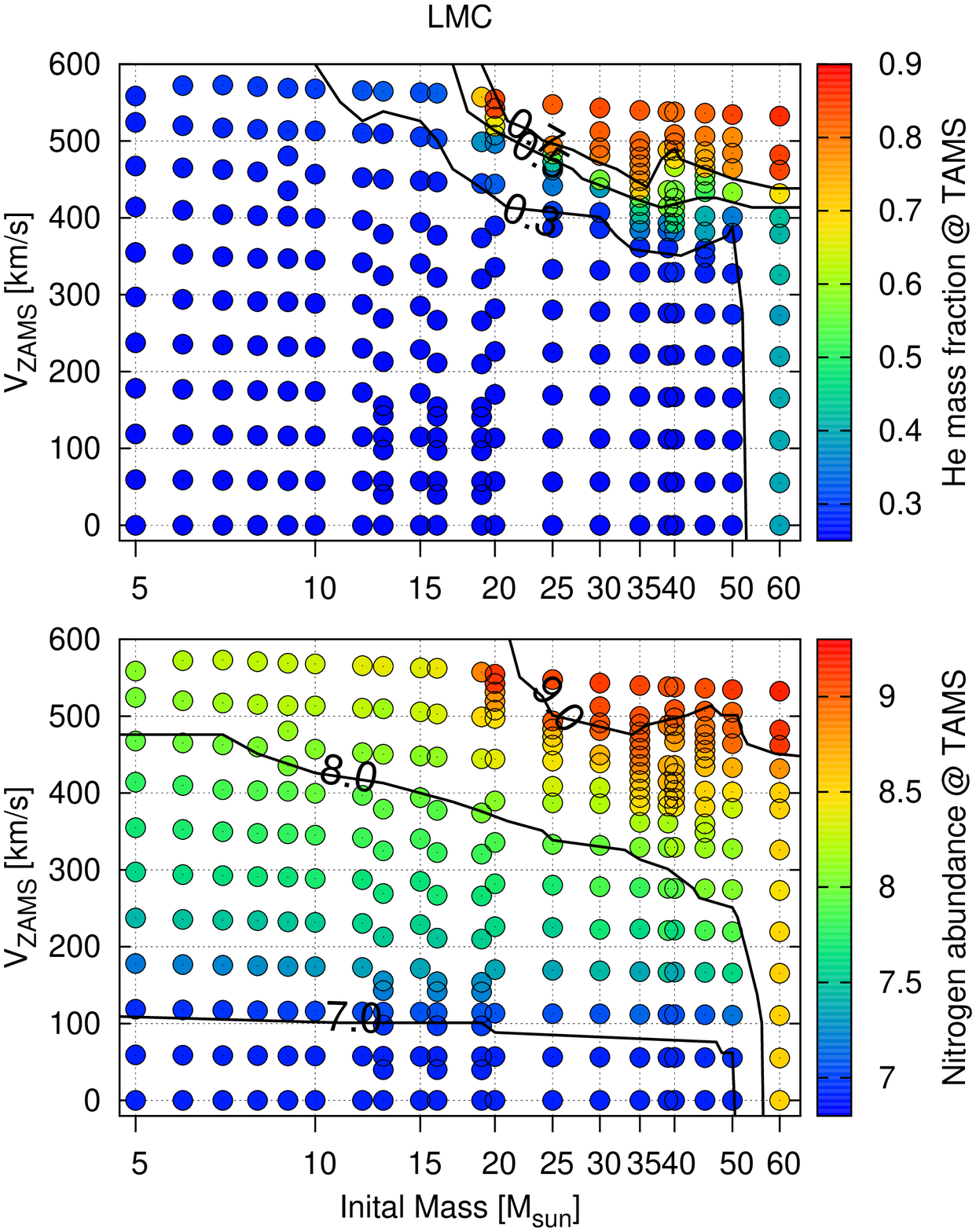}%
\includegraphics[bb=3 25 480 732, clip, angle=0,width=.31\textwidth]{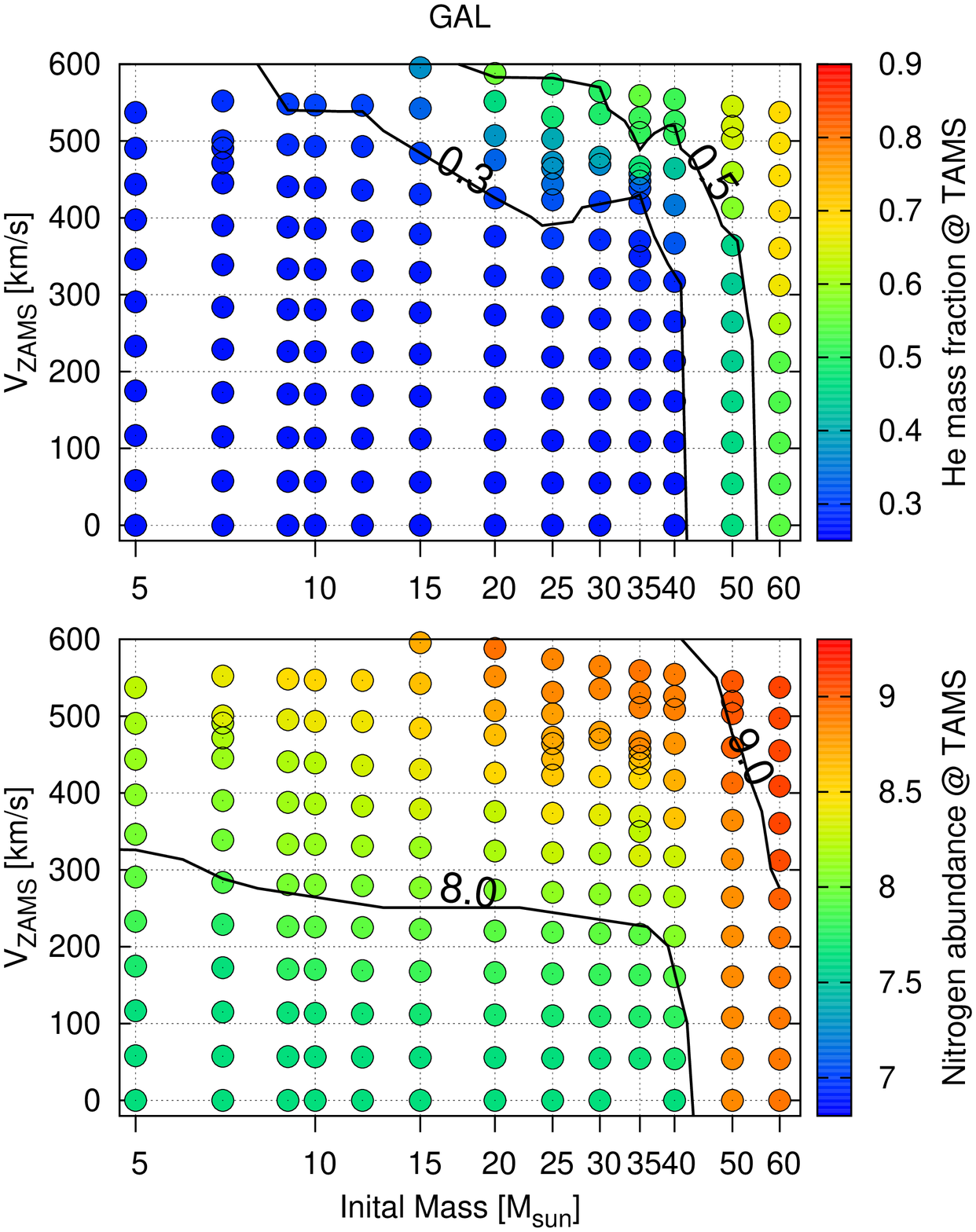}%
\includegraphics[bb=480 25 572 732, clip, angle=0,width=.06\textwidth]{16113fig4c.ps}

\caption{\label{fig:parameterspace}  Surface abundance of nitrogen (bottom row) and helium (top row) reached at the end of the main sequence (colors and contours) in our models with different initial masses, rotational velocities and compositions. Each evolutionary sequence of our grids is represented by one dot in this diagram. See Sec.~\ref{sec:init} for a description of the initial parameter space of our grid and Sec.~\ref{sec:abund} for a discussion of the abundances.}
\end{center}
\end{figure*}

\subsection{Grid of evolutionary tracks and isochrones: initial parameters and description of provided data}
\label{sec:init}
Using the calibration and the three initial compositional mixtures described above we computed a grid of evolutionary models. We adopt initial masses between 5 and 60\Msun and initial equatorial rotational velocities ($\vzams$) ranging from 0 to about $600\,\kms$. For each initial composition we computed about 200 evolutionary tracks.  The spacing between the initial parameters was chosen to provide a proper coverage of the parameter space and to provide extra resolution for massive fast rotating stars whose evolution depends sensitively on the initial rotation rate (see Fig.~\ref{fig:parameterspace}).

The results of our calculations are made available online through VizieR in table format.  For each evolutionary track we provide the basic stellar parameters at different ages such as the mass, effective temperature, luminosity, radius and  mass-loss rate as well as the surface abundances of various elements. We also list the central and surface mass fractions of the most important isotopes of these elements.  In Table~\ref{tab:trackdata} we summarized all the provided data.  The time steps are chosen to resolve the evolutionary changes of the main stellar parameters. Tracks with higher time resolution are available from the authors upon request.

Based on these evolutionary tracks we generated sets of isochrones, using our population synthesis code STARMAKER \citepalias{Brott10_popsyn}.  The isochrones cover the evolution until central hydrogen exhaustion for stellar masses between 5 and 60\Msun.  We provide them for ages between 0 and 35\,Myr at intervals of 0.2\,Myr, for initial rotational velocities between 0 and 540\,\kms with steps of 10\,\kms and for three different initial compositions: SMC, LMC and a Galactic mixture.  We provide the basic stellar parameters and surface abundances as listed in Tab.~\ref{tab:aiso_data}.

\begin{table}[htdp]
\caption{Description of the online tables containing the results of our evolutionary calculations as a function of time, for given initial composition, initial mass and initial rotation rate.}
\begin{center}
\begin{tabular}{l|c|p{0.5cm}p{0.5cm}p{0.5cm}}
\hline\hline
stellar  &       surf.           &\multicolumn{3}{c}{surface \& core } \\
parameters &  abun  &\multicolumn{3}{c}{isotope mass fractions}  \\
\hline
Age (t [yrs]) & H & $^1$H   & & \\
Mass (m [\Msun]) & He & $^3$He & $^4$He &\\ 
Eff. temp. ($\Teff$[K])& Li &  $^7$Li && \\
Luminosity ($\log \rm(L/L_{\odot})$)& Be & $^9$Be & & \\
Radius (R [$\rm{R_{\odot}}$])&B  &  $^{10}$B &  $^{11}$B & \\
Mass loss rate (\.{M} [\Msun/yr])& C & $^{12}$C & $^{13}$C \\
Surface gravity ($\logg$) & N & $^{14}$N & $^{15}$N &  \\
Surface velocity ($\varv_{\rm{surf}}$ [\kms]) & O & $^{16}$O & $^{17}$O &$^{18}$O  \\
Rotation period (P [days])& F & $^{19}$F & &  \\
Critical velocity  ($\varv_{\rm{crit}}$ [\kms]) & Ne & $^{20}$Ne &  $^{21}$Ne&  $^{22}$Ne \\
Eddington factor ($\Gamma_{\rm e}$)& Na&  $^{23}$Na& & \\
 & Mg &  $^{24}$Mg & $^{25}$Mg &$^{26}$Mg  \\
 & Al & $^{26}$Al &$^{27}$Al &  \\
 & Si & $^{28}$Si &$^{29}$Si &$^{30}$Si  \\
 & Fe & $^{56}$Fe & &  \\
 \hline
\end{tabular}
\tablefoot{The critical velocity is computed as $\varv^2_{\rm{crit}}$= $GM/R(1-\Gamma_{\rm e})$, where  $\Gamma_{\rm e}$ denotes the Eddington factor due to electron scattering.}
\end{center}
\label{tab:trackdata}
\end{table}%

\begin{table}[htdp]
\caption{Description of the online tables containing stellar parameters along isochrones computed for different initial rotational velocities
as provided by our population synthesis code \citepalias{Brott10_popsyn}.
See also Tab.~\ref{tab:trackdata}.}
\begin{center}
\begin{tabular}{l|c}
\hline\hline
stellar  &       surf.       \\
parameters &  abun  \\
\hline
Age (t [yrs]) & H \\
Mass (m [\Msun]) & He\\ 
Effective temperature ($\Teff$[K])& B \\
Luminosity ($\log \rm(L/L_{\odot})$)& C \\
Radius (R [$\rm{R_{\odot}}$])& N\\
Mass loss rate (\.{M} [\Msun/yr])& O \\
Surface gravity ($\logg$) & Ne\\
Surface velocity ($\varv_{\rm{surf}}$ [\kms]) & Na\\
Rotation period (P [days])& Mg   \\
Critical velocity  ($\varv_{\rm{crit}}$ [\kms]) & Si\\
Eddington factor ($\Gamma_{\rm e}$)& $^{1}$H (mass fraction) \\
initial mass and velocity & $^{4}$He (mass fraction)\\
  \hline
\end{tabular}
\end{center}
\label{tab:aiso_data}
\end{table}

\section{Evolutionary tracks  and isochrones}
\label{sec:models_hrd}
\label{sec:evolgrid}

\subsection {Evolutionary tracks}

Rotation affects evolutionary tracks in the Hertzsprung-Russell diagram \citep[e.g.][and references therein]{MaederMeynet2000_Review}.   The centrifugal acceleration reduces the effective gravity resulting in cooler and slightly less luminous stars. However,  rotation also induces internal mixing processes which can have the opposite effect, leading to more luminous and hotter stars.  Which of these effects dominates depends on the initial mass, rotation rate, metallicity and the evolutionary stage.  In this section we describe the effects of rotation on our models.

In rotating stars, the radiative energy flux depends on the local effective gravity \citep{vonZ24}, 
and temperature and luminosity become latitude dependent. 
The resulting thermal imbalance drives large scale meridional currents. 
In most stars, rotating at moderate rotation rates, the expected latitude dependence of temperature 
and luminosity is week \citep{Hunter09_nitrogen}. 
In models close to critical rotation, luminosity and temperature differences 
are expected to be more significant and may even give rise to polar winds and equatorial 
outflows caused by critical rotation \citep{Maeder99}. 
Temperatures, luminosities and gravities given for the models presented in this paper are surface averaged values.

\begin{figure}[!t]
\centering
\includegraphics[angle=-90,width=0.5\textwidth]{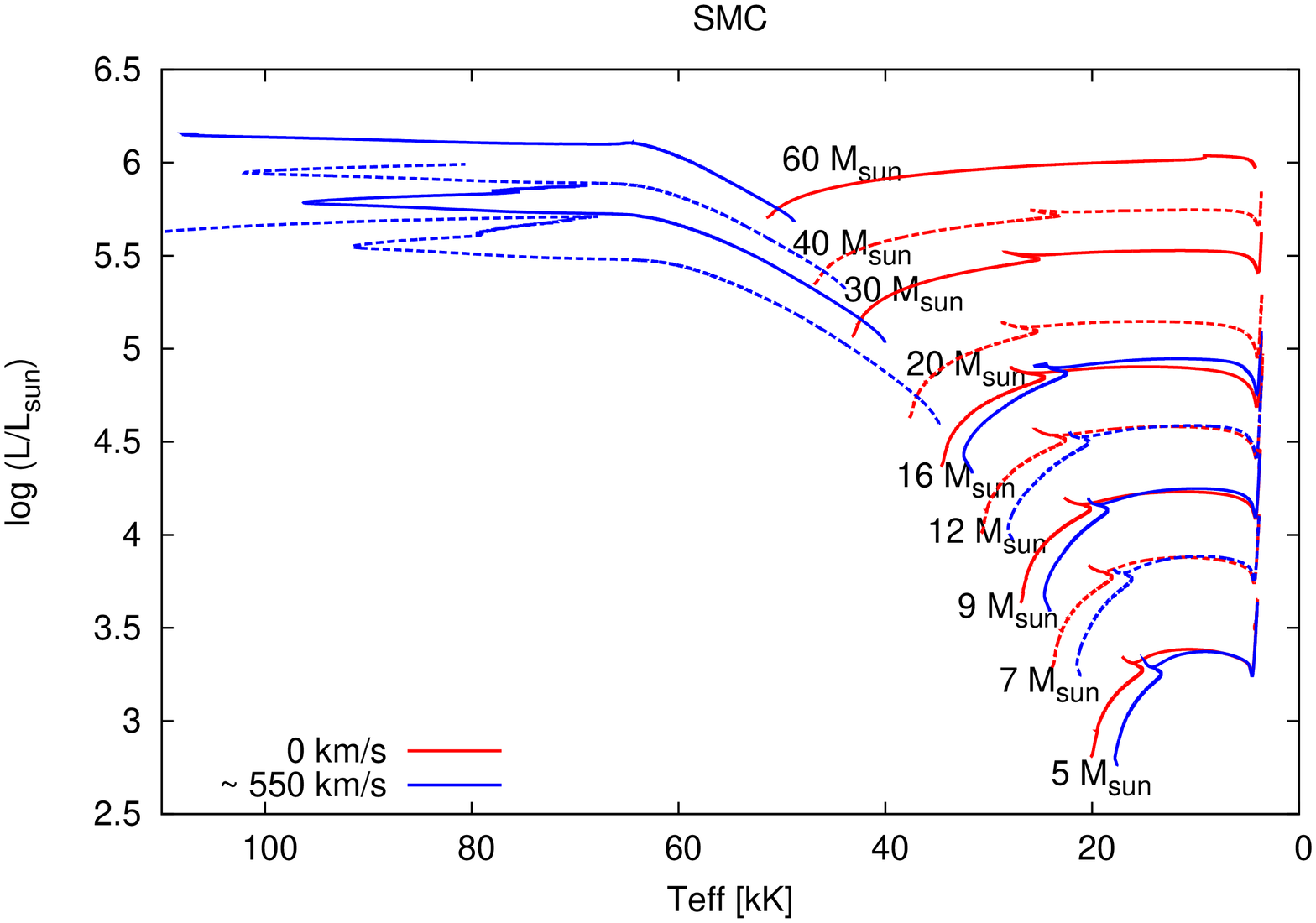}
\includegraphics[angle=-90,width=0.5\textwidth]{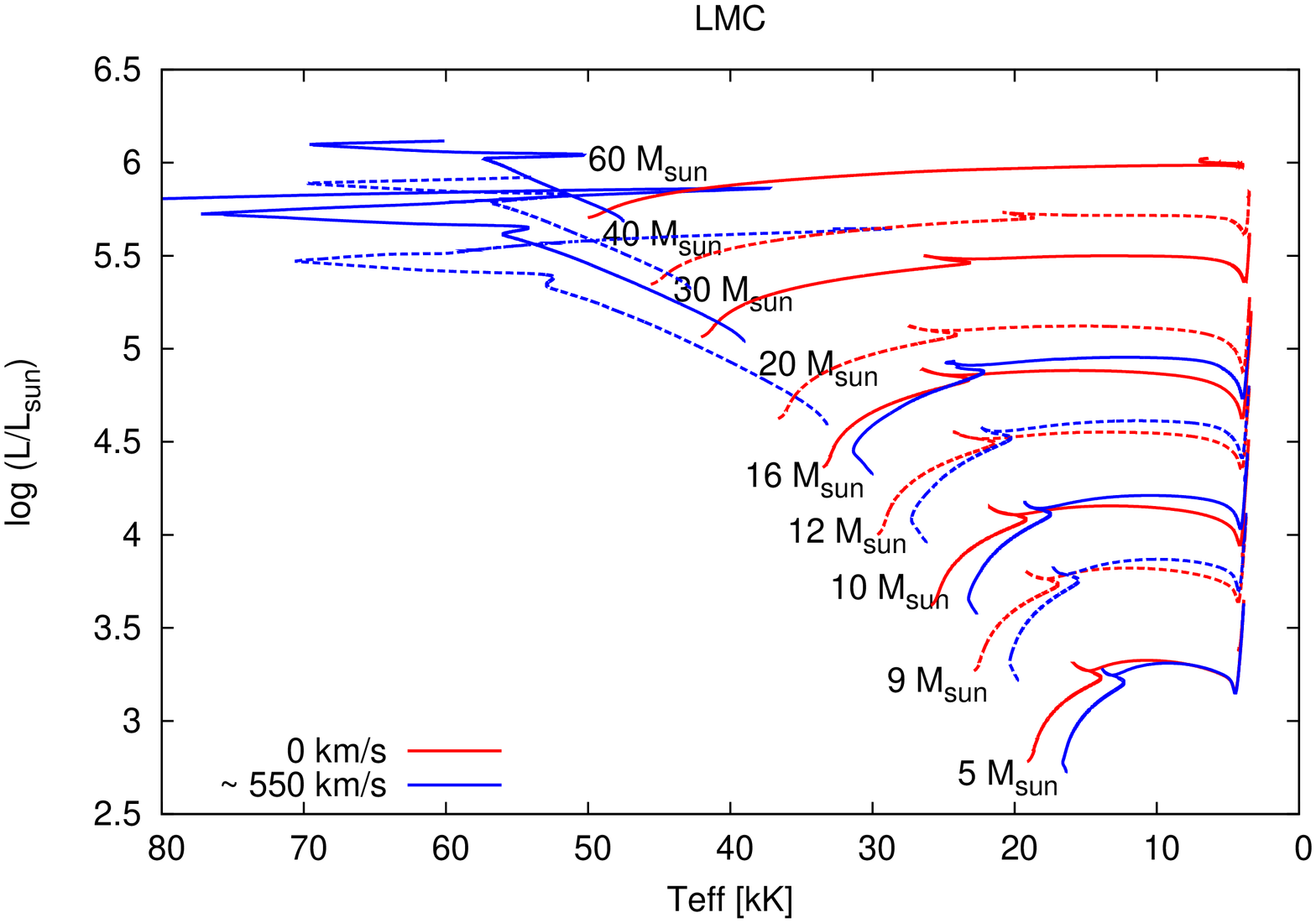}
\includegraphics[angle=-90,width=0.5\textwidth]{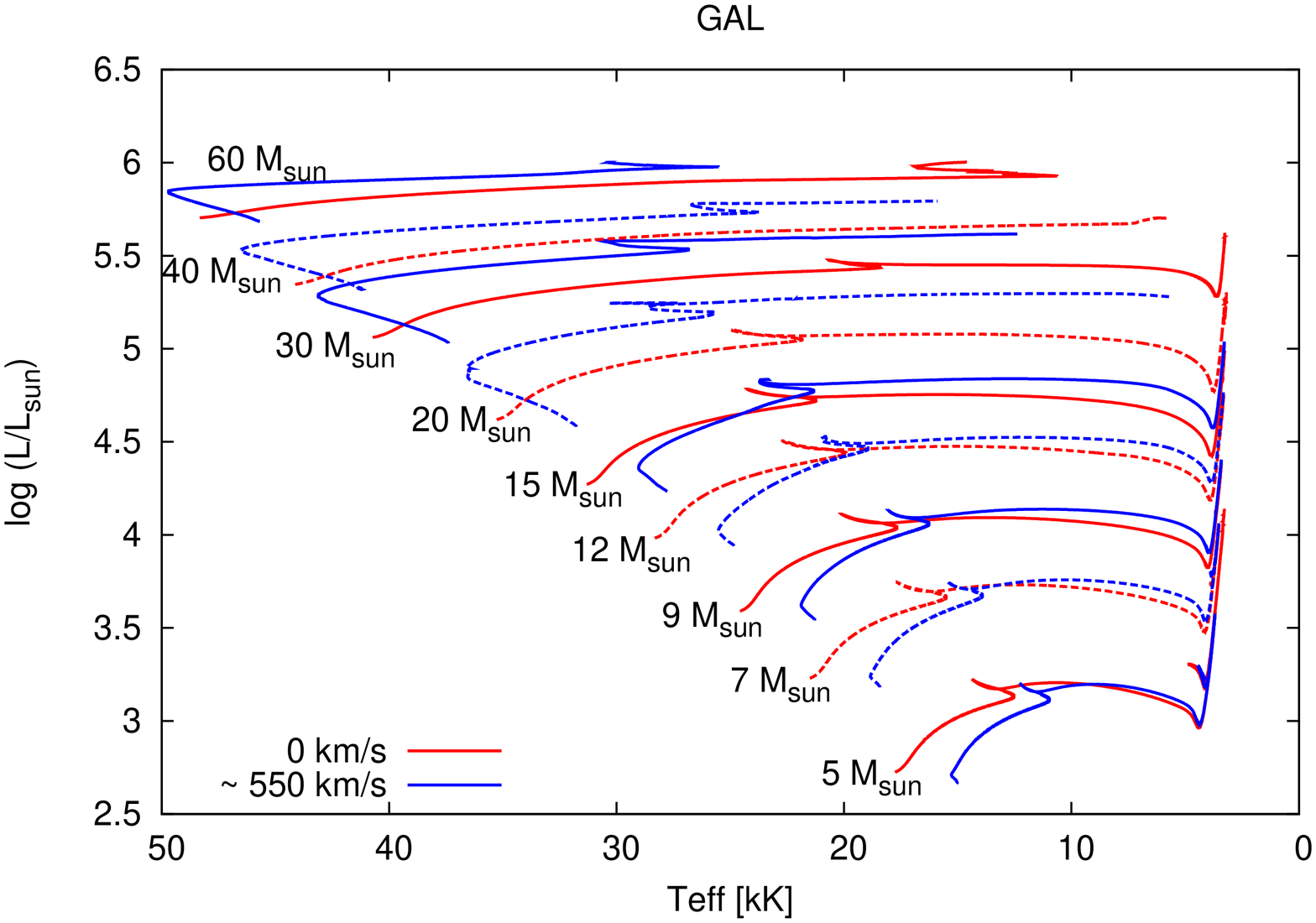}
\caption{Evolutionary tracks for various masses and metallicities. Red tracks are the non-rotating models and in blue we show models rotating initially at about 550\,\kms. For clarity, we alternate different line styles and scaled the temperature ranges such that the maximum of each plot is visible.}
\label{fig:hrd_tracks_mvar_vconst}
\end{figure}

\begin{figure}[!t]
\centering
\includegraphics[angle=-90,width=0.5\textwidth]{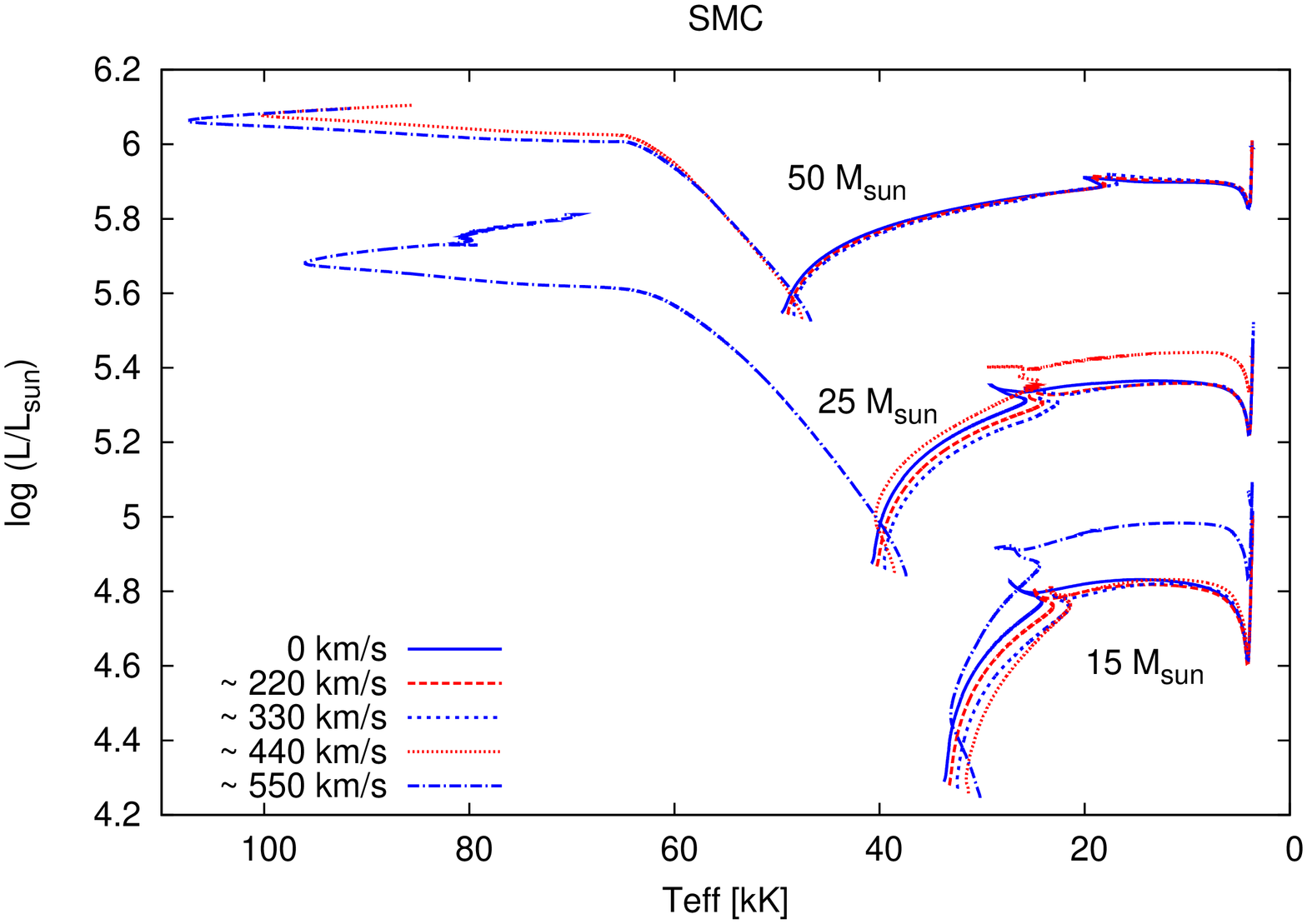}
\includegraphics[angle=-90,width=0.5\textwidth]{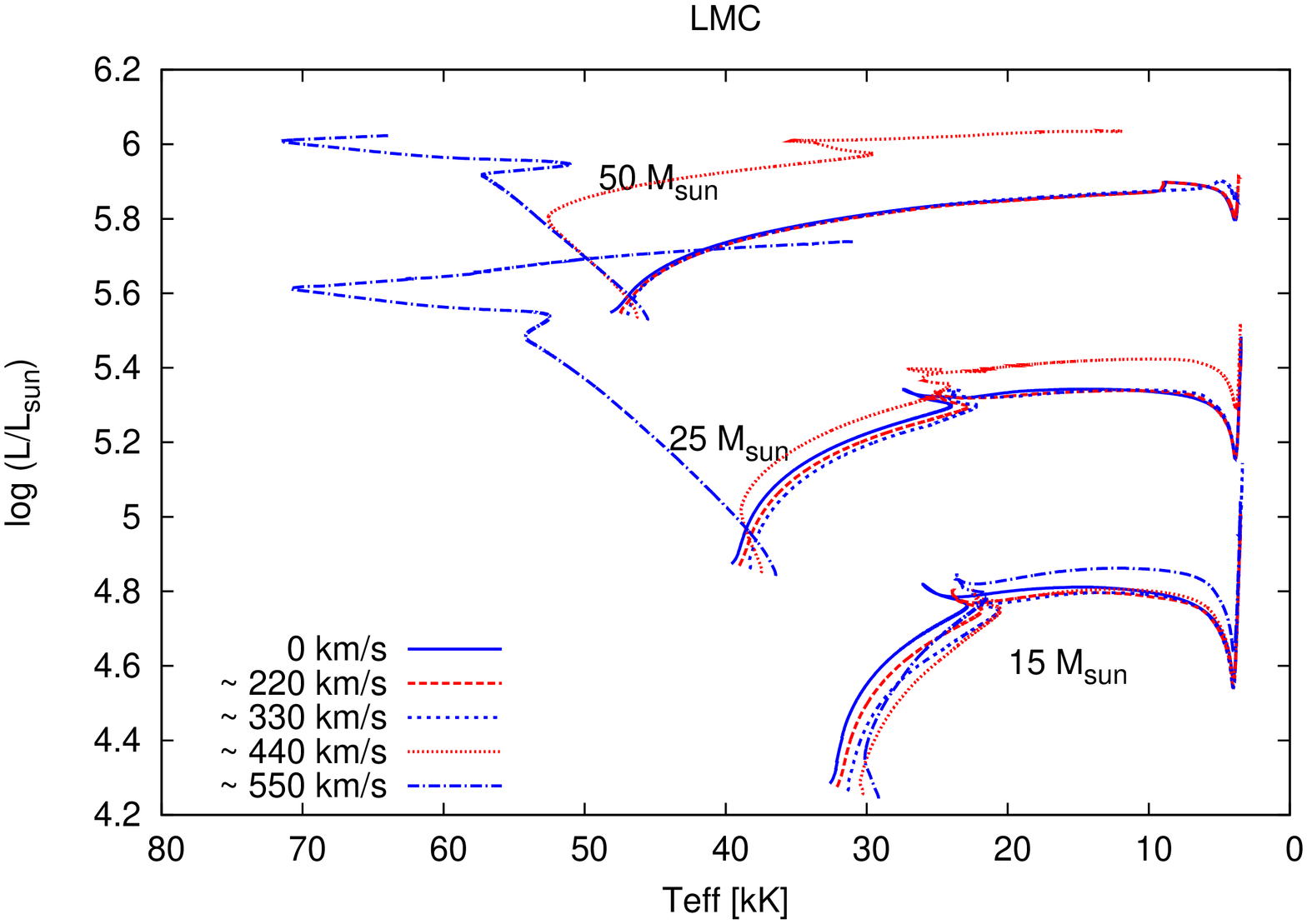}
\includegraphics[angle=-90,width=0.5\textwidth]{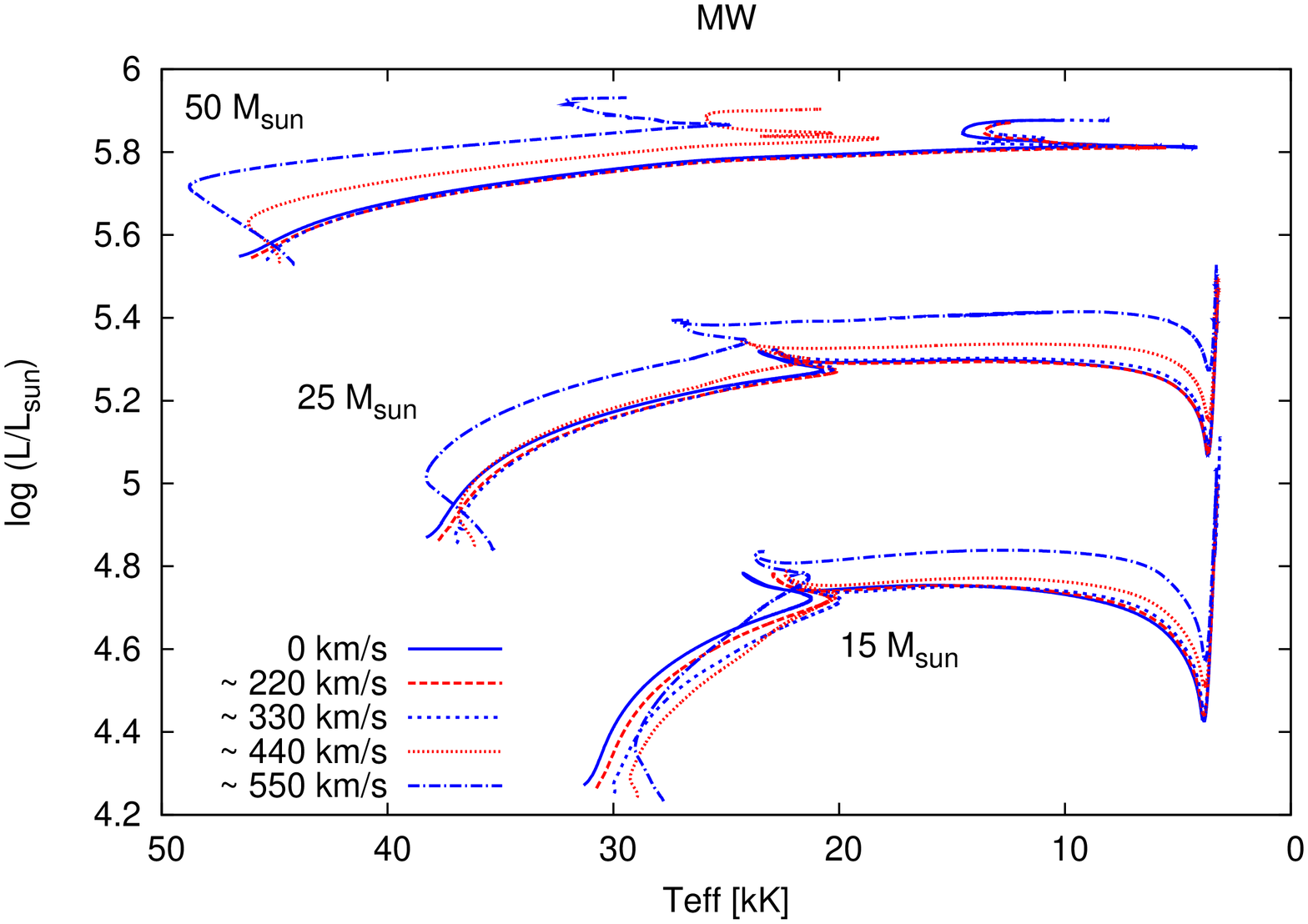}
\caption{Evolutionary tracks for 15, 25 and 50\Msun for SMC, LMC and Galactic composition (from top to bottom),
for a range of initial rotational velocities (see legend). }
\label{fig:hrd_tracks_mconst_vvar}
\end{figure}

\paragraph{Mass dependence of rotational mixing:}

For intermediate mass main-sequence stars the main effect of rotation is a reduction of the effective gravity by centrifugal acceleration.  
Fig.~\ref{fig:hrd_tracks_mvar_vconst} shows that, during the main-sequence evolution, the tracks of the fast rotating 5~$M_\odot$ stars are about 2000\,K cooler compared to the nonrotating counterparts.
Rotational mixing does affect the surface abundances of those elements
that are so fragile that the relatively low temperatures in the stellar envelope are
sufficient to induce nuclear reactions on them.
However, the mixing is not efficient enough to significantly alter the structure of the star. 
As a consequence, the effect of rotation on the corresponding evolutionary tracks in the Hertzsprung-Russell diagram remains limited. 

In rapid rotators more massive than about 15\Msun, the effect of rotational mixing becomes more important 
than that of the reduced effective gravity: towards the cool edge of the main sequence band, the rotating stars 
become more luminous than the nonrotating ones. 
The main effect of rotation is an effective increase of the size of the stellar core, similar to the effect of 
overshooting (Sec.~\ref{sec:overshooting}). This can be seen in Fig.~\ref{fig:hrd_tracks_mvar_vconst} when one compares 
the tracks of the 15 and 16\Msun models near the end of their main-sequence evolution. 
The larger core mass in the fast rotating models results in a higher luminosity. %

In the most massive stars at low metallicity ($ \gtrsim 16$\Msun for the SMC, $M \gtrsim 19$\Msun for the LMC), and for the extreme rotators shown in Fig.~\ref{fig:hrd_tracks_mvar_vconst}, mixing induced by rotation is so efficient that almost all the helium produced in the center is transported throughout the entire envelope of the star.   During their main-sequence evolution they become brighter and hotter, evolving up- and bluewards in the Hertzsprung-Russell diagram.  After all hydrogen in the center has been converted to helium, the star contracts, reaching effective temperatures of up to 100,000\,K.  This type of evolution is referred to as ``quasi-chemically homogeneous evolution'' \citep{Maeder87, Yoon05, Woo06}. The blueward and redward evolution after the exhaustion of hydrogen in the center is analogous to the "hook" at the end of the main sequence seen in tracks of nonrotating stars. Note that all models have been computed beyond central hydrogen exhaustion, in most cases up to the ignition of helium.
This allows us to later compare the main sequence surface abundances with those resulting from the first dredge-up
in the red supergiant regime.

\paragraph{Metallicity dependence of rotational mixing: }

Fig.~\ref{fig:hrd_tracks_mconst_vvar} depicts evolutionary tracks of models with various initial rotation rates. The most striking feature in these diagrams is the bifurcation of the evolutionary tracks occurring at high masses and low metallicity, most clearly visible at SMC metallicity.   Stars that rotate faster than a certain threshold are so efficiently mixed that they evolve almost chemically homogeneously.  Stars that rotate slower than this threshold build a chemical gradient at the boundary between the convective core and the radiative envelope. This gradient itself has an inhibiting effect on the mixing processes, strongly reducing the transport of material from the core to the envelope.  The minimum rotation rate required for chemically homogeneous evolution decreases with increasing mass \citep{Yoon06}. 

At high metallicity, rotational mixing is less efficient. In addition, mass and angular momentum loss due to stellar winds becomes important, slowing down the rotation rate and therefore the efficiency of the mixing processes.  The fastest rotating stars at high metallicity initially evolve blue- and upward in the Hertzsprung-Russell diagram, see in the lower panel of Fig.~\ref{fig:hrd_tracks_mconst_vvar}.  However, the combined effects of spin down by a stellar wind and the build up of an internal chemical gradient reduces the efficiency of internal mixing processes.  The star switches onto a redward evolutionary track, similar to that of a non rotating star. However, its larger core mass results in a higher luminosity compared to the slower rotating counterparts.

\begin{figure*}[htpb]
\begin{center}
\includegraphics[width=0.33\textwidth,bb=140 0 450 800, clip]{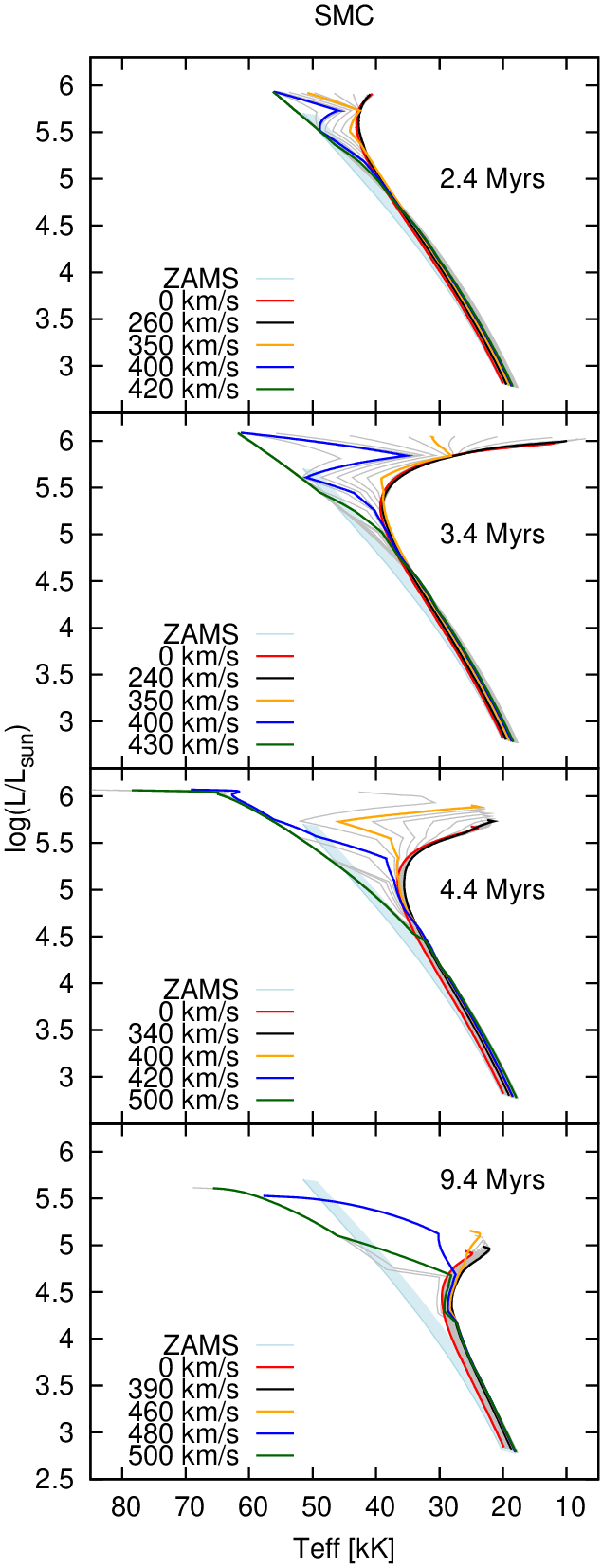}%
\includegraphics[width=0.33\textwidth,bb=140 0 450 800, clip]{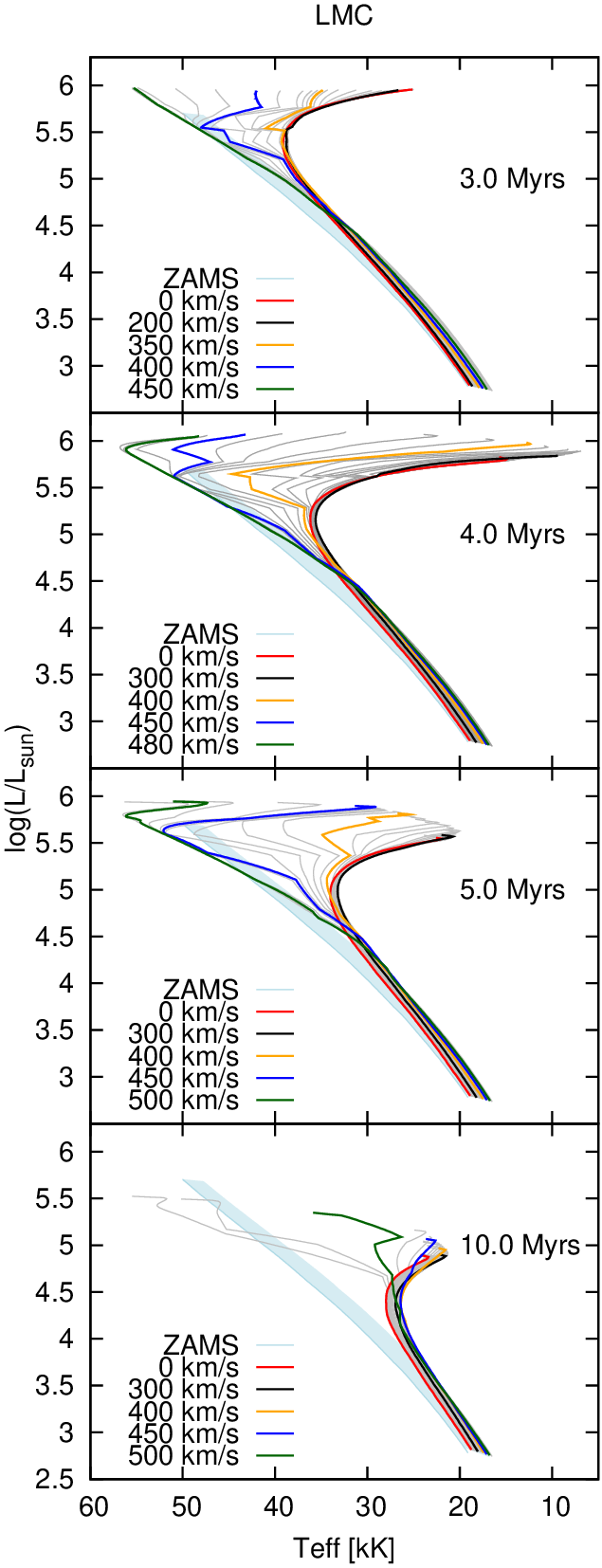}%
\includegraphics[width=0.33\textwidth,bb=140 0 450 800, clip]{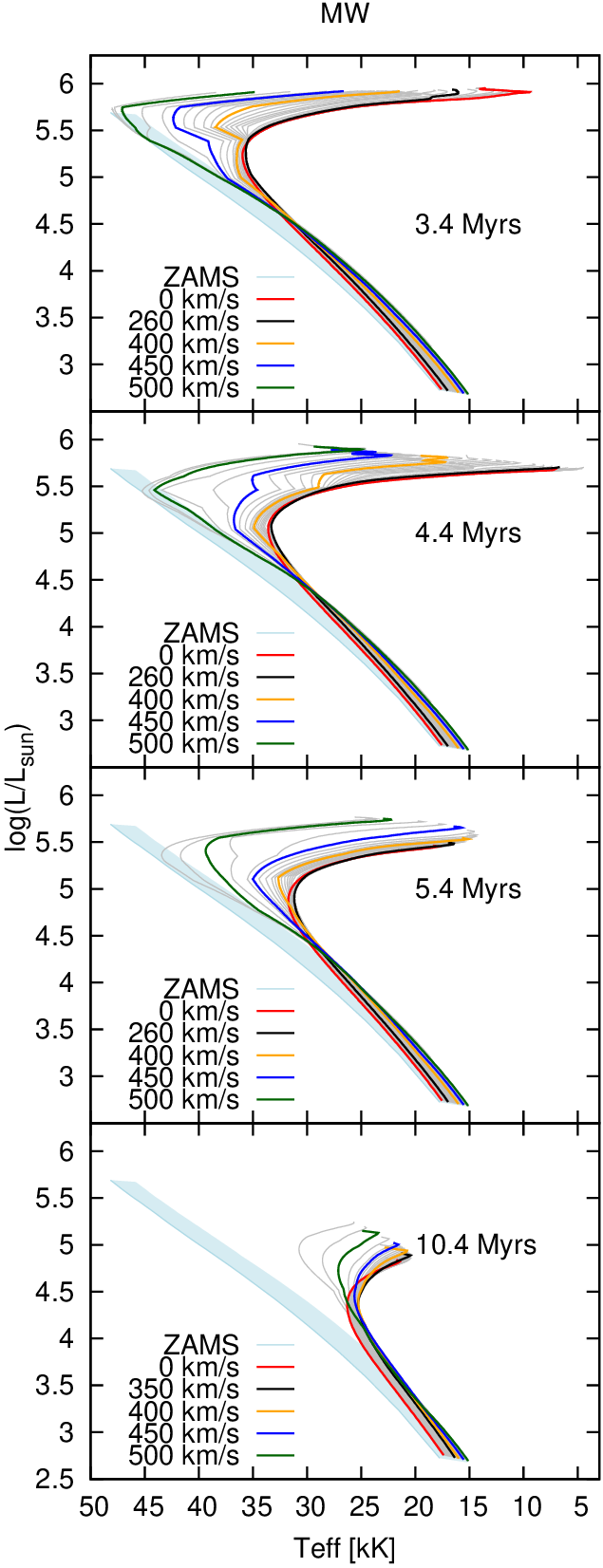}%
\caption{Isochrones for SMC (left panels), LMC (center) and galaxy (right) composition. Each panel shows isochrones for a given age and initial rotational velocity ranging from  0 to 540\,\kms in steps of 20\,\kms, with an increased resolution of 10\,\kms between 350 and 450\,\kms (grey lines). In red we show the isochrone for non rotating models, for other colors see the legend and the main text. We show the area spanned by the zero-age isochrones as a light blue band.  The age in the second row has been chosen such that the isochrones span the largest range in effective temperature, 1 Myr earlier is shown in the first row, 1 and 5 Myr later are shown in the third and fourth row. }
\label{fig:aiso}
\end{center}
\end{figure*}

\subsection{Isochrones}
\label{sec:aiso-hrd}
While for a given metallicity, a classical isochrone can be represented by a single line in the Hertzsprung-Russell diagram, the isochrones of rotating stars span an area for a given age and initial composition. This is shown in Fig.~\ref{fig:aiso}. The isochrone constructed from nonrotating evolutionary models is plotted in red. The isochrone corresponding to the fastest initial rotational velocity for which the models do not yet follow a blueward evolution in the HRD is plotted in black. The isochrone with the slowest initial velocity that shows a clear bluewards evolution is shown in green. In blue and yellow we have selected two isochrones from the transition region that may help the reader to asses the sudden transition from classical to chemically homogeneous evolution.  

For rotational velocities above 350\,\kms, when the most massive stars undergo chemically homogeneous evolution,  the isochrones deviate strongly from the nonrotating case. The maximum spread in effective temperature occurs around 4 Myr 
(second row in Fig.~\ref{fig:aiso}). 
At lower metallicity the  isochrones split into more clearly separated branches. At SMC metallicity (e.g. the third panel in the first column in Fig.~\ref{fig:aiso}) the isochrone based on models rotating initially at 500\,\kms moves straight to the blue.  In contrast, the comparable LMC isochrone returns to the red for stars above $\sim$50\Msun. This behavior is directly related to the feature in the evolutionary LMC tracks in Fig.~\ref{fig:hrd_tracks_mconst_vvar}, see for example the 50\Msun track for 550\,\kms.  At Galactic metallicity the blueward evolution does not appear in the models.  Nevertheless, the area spanned by the isochrones extends over a wide range of effective temperatures. 

Between about 5 and 10 Myr the most massive nonrotating stars have evolved off the main sequence. 
However, at low metallicity, the most massive fast rotators, which undergo chemically homogeneous evolution 
are still in their main-sequence phase at this time,  forming a blue straggler-like blue population (see the bottom panels of Figs.\ref{fig:aiso}).   If homogeneously evolving stars exist in nature they would most likely be found in low metallicity star clusters with ages between 5 and 10\,Myr.

%--------------------------------------
\section{Surface Abundances}
\label{sec:tracks_surf_changes}
\label{sec:abund}
\subsection {Abundances as a function of time}

\begin{figure*}[p]
\centering
\includegraphics[angle=0,width=\textwidth, bb= 0 160 600 800, clip]{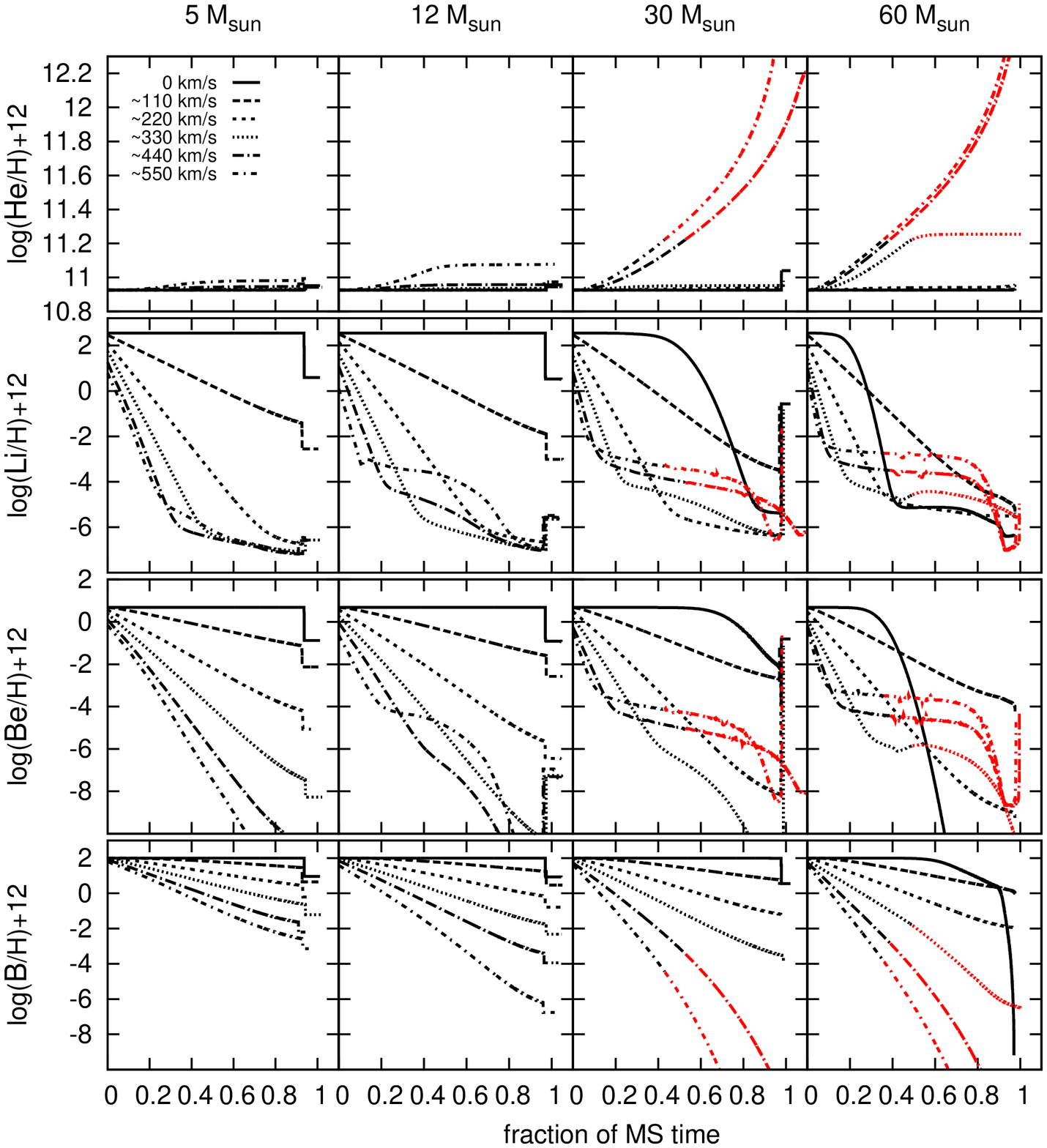}
\caption{Change of the surface abundances of helium, lithium, beryllium and boron along evolutionary tracks of SMC composition as a function of time, expressed as a fraction of the main-sequence lifetime. Different line styles represent different initial rotational velocities (see legend).  
For models where the surface helium mass fraction exceeded 40\%, the abundances are plotted in red, to indicate
that the reference element hydrogen is depleted significantly (see also Sec.~\ref{sec:tracks_surf_changes}).
}
\label{fig:track_abundance_changes1_smc}
\end{figure*}

A direct observable consequence of rotationally induced mixing is the enrichment or depletion of certain elements in the atmospheres of main-sequence stars.  In Fig.~\ref{fig:track_abundance_changes1_smc}-\ref{fig:track_abundance_changes1_mw} we show the evolution of the surface abundance of various elements as a function time.  The effects of rotational mixing are more pronounced at lower metallicity, at higher masses and for
 higher rotational velocity (see also Fig.~\ref{fig:parameterspace} and Sec.~\ref{sec:evolgrid}).  

As is usual in observational work, we express the surface abundances relative to the abundance of hydrogen. When stars become significantly hydrogen depleted at the surface, using hydrogen as a reference element may not be the most logical choice. Changes in the abundance may partially reflect changes in the reference element hydrogen.  We plot the abundances in red when these effects become important (i.e. when the helium mass fraction at the surface becomes larger than 40\%) .

Most of our stellar models evolve to the red supergiant stage directly after the end of core hydrogen burning. 
This leads to a large vertical step in the surface abundances of many elements 
in Fig.~\ref{fig:track_abundance_changes1_smc}-\ref{fig:track_abundance_changes1_mw},
which is due to the convective dredge-up in the red supergiant stage.
In the following, we discuss the evolution of the surface abundances of various groups of elements, focusing on the changes occurring over the course of the main-sequence evolution.  

\subsubsection{Helium}
Even though helium is the main product of hydrogen burning, the abundance of helium at the surface remains remarkably constant in most evolutionary tracks during the main-sequence phase.  For the 12\Msun models the enhancement is less than 0.2 dex. Only the fast rotators of 30 and 60 \Msun show significant helium surface enrichments, especially at low metallicity (Fig.~\ref{fig:track_abundance_changes1_smc}-\ref{fig:track_abundance_changes1_mw}). These behaviors can be understood as the combination of two effects. Firstly, the production of helium occurs on the nuclear timescale. This occurs slower than, for example, the production of nitrogen or destruction of Li.  Secondly, with the production of helium a steep gradient in mean molecular weight is established at the boundary between the core and the envelope.  Such gradients inhibit the efficiency of mixing processes and prevent the transport of helium to the surface.    
Significant amounts of helium can be transported to the surface in models were mixing processes are efficient enough to prevent the build-up of a chemical barrier between core and envelope.

\subsubsection{The fragile elements, Li, Be, B and F}
\label{sec:LiBeB}
The elements Li, Be, B are destroyed by proton captures at temperatures higher than $2.5\times 10^{6}\,$K for lithium, $3.5\times 10^{6}\,$K  for beryllium, $5 \times 10^{6}\,$K for boron \citep[][p. 123]{McWilliam04}.  These elements can only survive in the outermost layers.  In rotating stars the surface abundances of these elements decrease gradually as the outer layers are mixed with deeper layers in which these elements have been destroyed. The decrease is most rapid for the most fragile element, lithium. The surface abundance of fluorine behaves similarly. It can survive temperatures of up to about $20 \times 10^{6}$K. We note that the rates of the reactions in which fluorine is involved are quite uncertain \citep{Arnould99}. 

In addition, stellar winds can affect the surface abundances especially for the most massive stars at high metallicity.  Due to mass loss deeper layers are revealed in which the fragile elements have been destroyed.  For nonrotating stars a sudden drop in the surface abundances of the fragile elements is visible in Fig.~\ref{fig:track_abundance_changes1_smc}-\ref{fig:track_abundance_changes1_mw}, when mass loss has removed the  layers in which these elements can survive.  In rotating models the change is more gradual.

After the end of the main-sequence, dredge-up leads to a further decrease of the surface abundances of these elements. However, for some models with masses between 20 and 40\Msun we find that small amounts of Li and Be are produced in the hydrogen burning shell by the p-p chain.  
As a result of the interplay between a convective zone on top of the shell source and the convective envelope reaching into these layers after hydrogen exhaustion, we 
find that the surface abundances of these elements can be significantly increased in these models.
Even though the possible production of these elements by massive stars has several interesting applications, the robustness of the predictions for these elements requires  further investigation.

\subsubsection {Carbon, nitrogen, oxygen and sodium}

Carbon, nitrogen and oxygen take part as catalysts in the conversion of hydrogen into helium.  Although their sum remains roughly constant, nitrogen is produced when the cycle settles into equilibrium, at the expense of carbon and, somewhat later, oxygen. 
The CN-equilibrium is achieved very early in the evolution, before a strong mean molecular weight gradient has been established between the core and the envelope. Therefore, rotational mixing can transport nitrogen throughout the envelope to the stellar surface. 
While the surface abundance of nitrogen increases, carbon is depleted. 
Since full CNO-equilibrium is achieved only after a significant amount of hydrogen has been burnt in the core,
changes in the oxygen surface abundance are only found in later stages in the more massive and faster rotating models. 

Sodium is produced in the extension of the CNO-cycle, the NeNa-chain.  The changes in the surface abundances of sodium resemble the changes is nitrogen, even though the enhancements are smaller and appear a little bit later at the stellar surface, see Fig.~\ref{fig:track_abundance_changes1_smc}-\ref{fig:track_abundance_changes1_mw}. 

\begin{figure}[htbp]
\begin{center}
\includegraphics[angle=-90, width=0.5\textwidth, bb=95 50 554 770, clip]{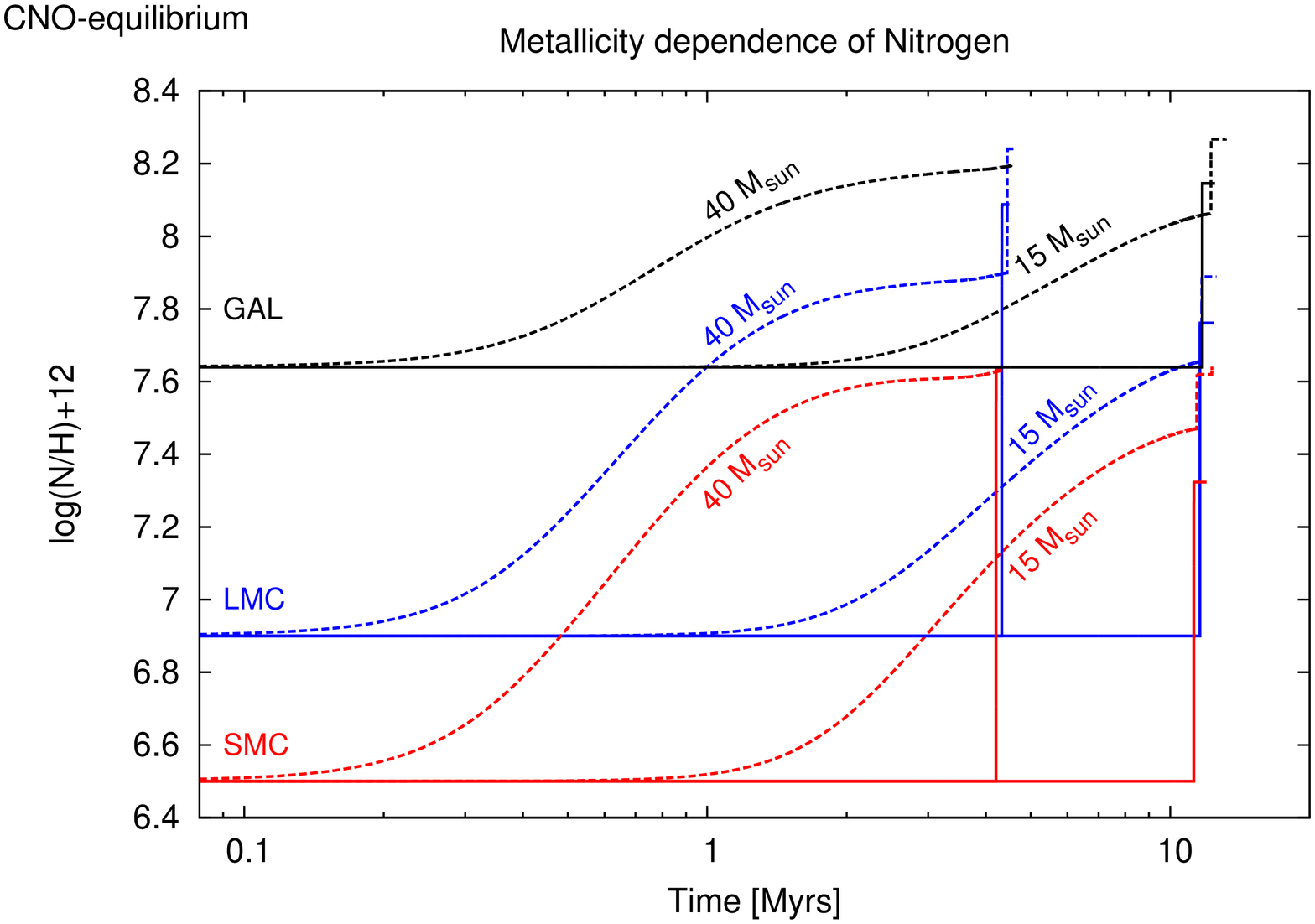}
\caption{Nitrogen as a function of time at SMC, LMC and Galactic metallicity. The models are of  15 and 40\Msun. Full lines represent  nonrotating models, dashed lines models rotating initially at $\sim$270\,\kms.} 
\label{fig:nitrogen-time-z}
\end{center}
\end{figure}

The relative increase of the nitrogen abundance depends on the initial amount of carbon available for conversion into nitrogen.  These amounts are different for the different mixtures. The C/N ratio in the SMC and LMC composition are similar (7.4 and 7.1, respectively), while the ratio in our Galactic composition is smaller (3.1).  In Fig.~\ref{fig:nitrogen-time-z} we show the evolution of the surface nitrogen abundance in models of 15 and 40\Msun with initial rotation rates of 0 and 270\,\kms for SMC, LMC and Galactic composition. The initial abundance of nitrogen in the Magellanic clouds is much lower than in the Galactic mixture. Nevertheless, the rotating SMC and LMC models reach surface nitrogen abundances at the TAMS which can be higher than in the initial Galactic mixture. The steep rise at the end of the curves is due to the dredge-up when the star becomes a red supergiant. The TAMS is located at the base of the steep rise.

\subsection{Surface abundances for selected ages}
\label{sec:aiso-abun}

Due to the effects of rotation, the age or mass of a star can no longer be determined uniquely 
from its position in the Hertzsprung-Russell diagram by fitting evolutionary tracks or isochrones. This holds especially for stars that are located near the zero-age main sequence. A young, unevolved, slowly rotating star may have the same effective temperature and luminosity as a less massive, fast rotating, evolved star.      A determination of the surface composition and to some extent the projected rotational velocity may help to lift this degeneracy. 

In the panels of Fig.~\ref{fig:aiso_abundance_changes_smc}-\ref{fig:aiso_abundance_changes_mw} we show
which abundance distributions are predicted at a given time, as function of the considered mass
and rotational velocity.
Models of more  massive stars show more pronounced surfaces abundances changes for a given age. 
The kinks in the lines of 400\,\kms at LMC and Galactic composition occur due to difficulties in the interpolation 
between stars that follow blueward and redward evolutionary tracks effects. 

These plots also show the effect of rotation on the main sequence life time. For example, Fig.~\ref{fig:aiso_abundance_changes_mw} shows that the lines representing models with initial rotation rates of 400 and 500\,\kms predict that stars of 50 and 60\Msun are still present  at 4\,Myr, whereas nonrotating models predict that they have left the main sequence.  On a much smaller scale this effect is also visible if one compares lines for rotation rates between 0 and 300\,\kms based on models that follow normal evolutionary tracks.    

The steep drop in the boron abundance along the isochrones based on nonrotating models is related to the stellar wind mass loss.  For example, at LMC composition,  the drop occurs around 35\Msun, indicating that the winds efficiently remove the outer layers of stars of 35 \Msun and higher
(see also Sec.~\ref{sec:LiBeB}). 
 
%:------------------------- Summary ----------------------------------------
\section{Summary}
\label{sec:summary}
In this paper we have presented an extensive grid of models for rotating massive main sequence stars.  We provide three sets of initial compositions that are suitable for comparison with early OB stars in the SMC, the LMC and the Galaxy. The models cover the main-sequence evolution and in most cases these have been computed up to helium ignition. 
In terms of overshooting and rotational mixing efficiency,
our models have been calibrated against the FLAMES survey of massive stars. 
We are using a new method to calibrate the amount of overshooting that makes use of the observed drop 
in projected rotation rates for stars with surface gravities lower than $\log g = 3.2$\, dex.  
Interpreting this drop as the end of the main sequence, we find an overshooting parameter of $0.34\pm 0.1$ pressure scale heights.

We have also presented a detailed set of isochrones based on models of rotating massive main sequence stars.  
Whereas classical isochrones can be represented by a single line in the Hertzsprung-Russell diagram, 
the isochrones of rotating stars span a wide range of effective temperatures at a given luminosity. 
Therefore, the mass and age of an observed star can no longer be determined uniquely from its location in the Hertzsprung-Russell diagram.  
We also provided detailed predictions for the changing surface abundances of rotating massive main sequence stars.
While we believe that the data provided here can be useful to many future studies of massive stars,
we make use of it in \citepalias{Brott10_popsyn} for undertaking a quantitative test of rotational mixing of massive
stars in the LMC. 

\acknowledgements{We thank Georges Meynet, the referee of this paper, for
many helpful comments and suggestions.

This work has been performed within the framework of the FLAMES consortium of massive stars and has made use of the VizieR catalogue access tool, CDS, Strasbourg, France. 
SdM acknowledges support for this work provided by NASA through Hubble Fellowship grant
HST-HF-51270.01-A awarded by the Space Telescope Science Institute,
which is operated by the Association of Universities for Research in
Astronomy, Inc., for NASA, under contract NAS 5-26555.
}
\bibliography{../../literatur}

%:---------- Appendix, plots of abundance changes along evolution tracks. 

\addtocounter{figure}{-1}

\begin{figure*}[p]
\centering
\includegraphics[angle=0,width=\textwidth, bb= 0 10 600 800, clip]{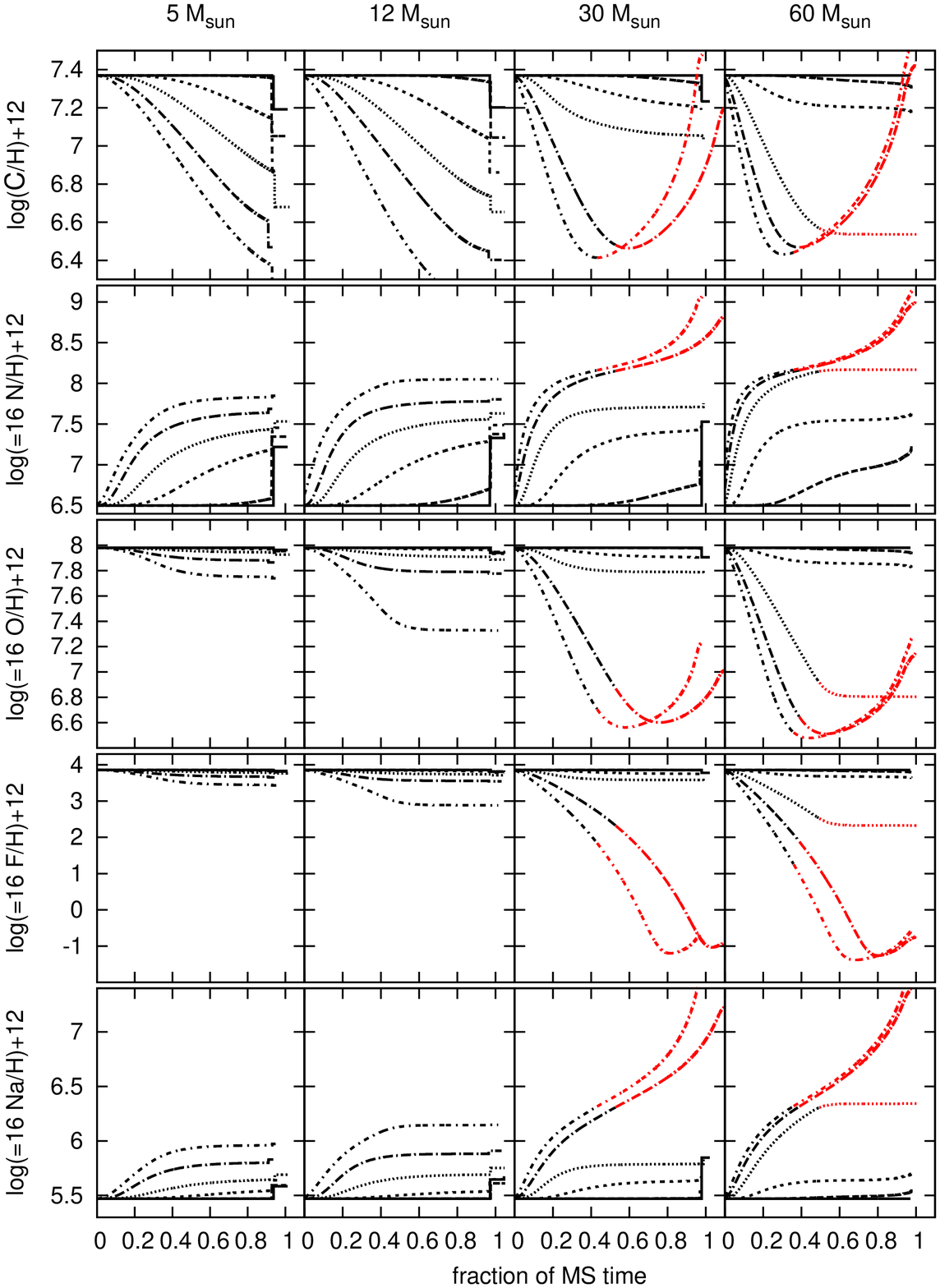}
\caption{Surface abundances of rotating models at SMC metallicity, continued. Shown are carbon, oxygen, nitrogen, fluorine and sodium (from top to bottom) as a function of time, expressed as a fraction of the main-sequence lifetime.}
\label{fig:track_abundance_changes2_smc}
\end{figure*}

%%---------- surface abundances lmc

\begin{figure*}[!t]
\centering
\includegraphics[angle=0,width=\textwidth, bb= 0 160 600 800, clip]{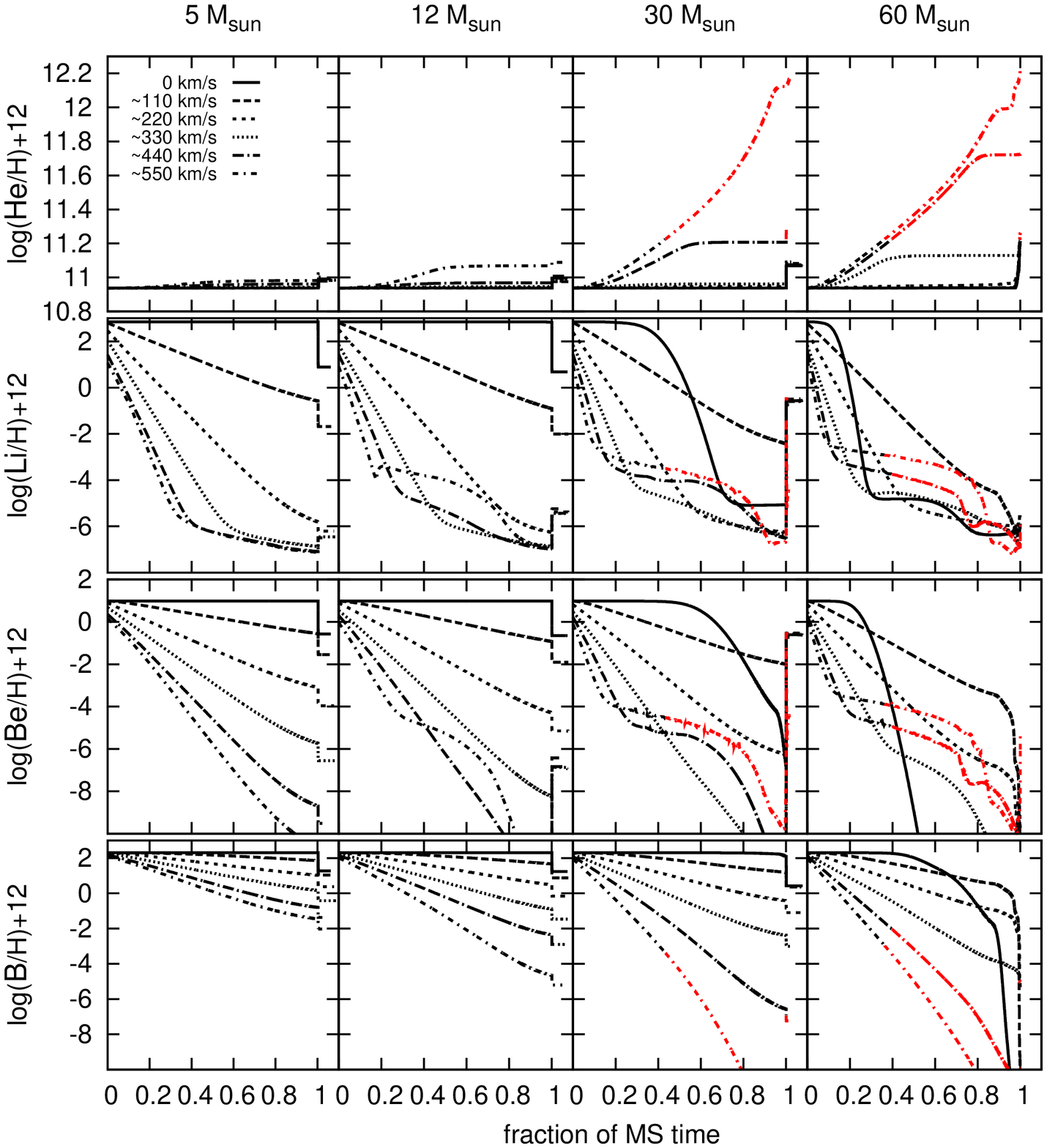}
\caption{Like Fig.~\ref{fig:track_abundance_changes1_smc}, for stellar models of LMC composition.}
\label{fig:track_abundance_changes1_lmc}
\end{figure*}

\addtocounter{figure}{-1}

\begin{figure*}[p]
\centering
\includegraphics[angle=0,width=\textwidth, bb= 0 10 600 800, clip]{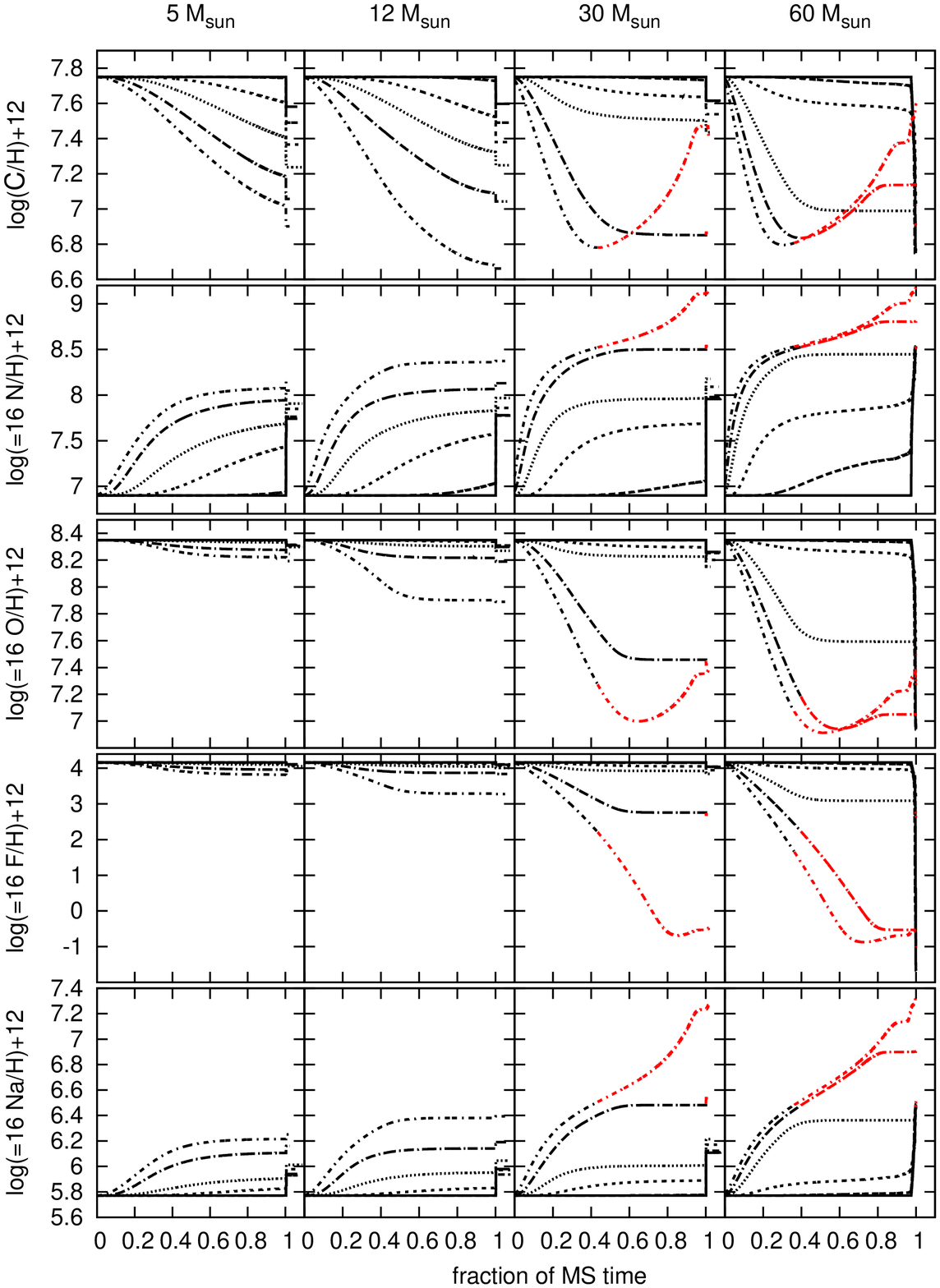}
\caption{continued: Like Fig.~\ref{fig:track_abundance_changes1_smc}, for stellar models of LMC composition.}
\label{fig:track_abundance_changes2_lmc}
\end{figure*}

%% surface abundances Galactic

\begin{figure*}[p]
\centering
\includegraphics[angle=0,width=\textwidth, bb= 0 160 600 800, clip]{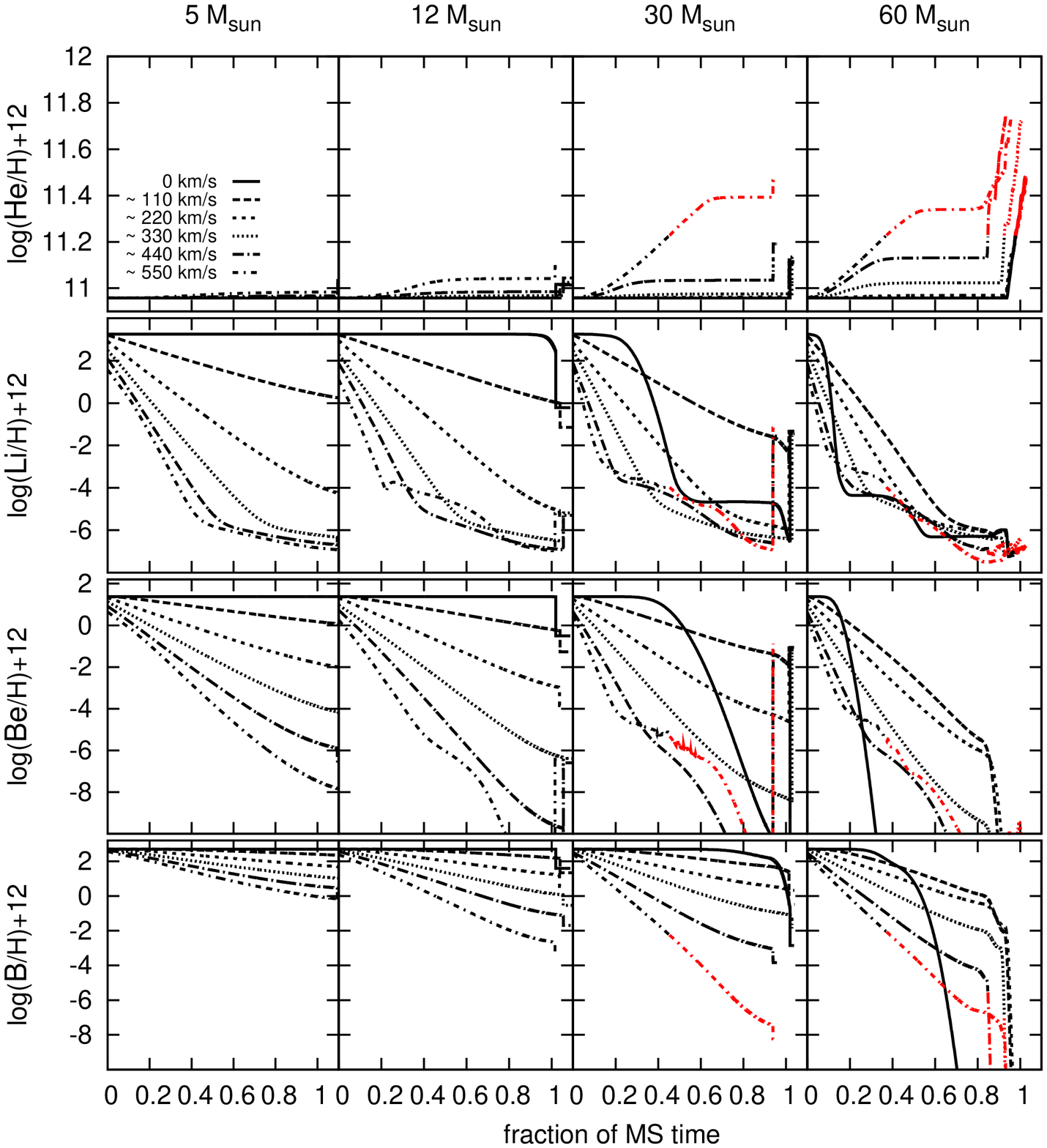}
\caption{Like Fig.~\ref{fig:track_abundance_changes1_smc} for stellar models with a Galactic initial composition.}
\label{fig:track_abundance_changes1_mw}
\end{figure*}
\addtocounter{figure}{-1}

\begin{figure*}[p]
\centering
\includegraphics[angle=0,width=\textwidth, bb= 0 10 600 800, clip]{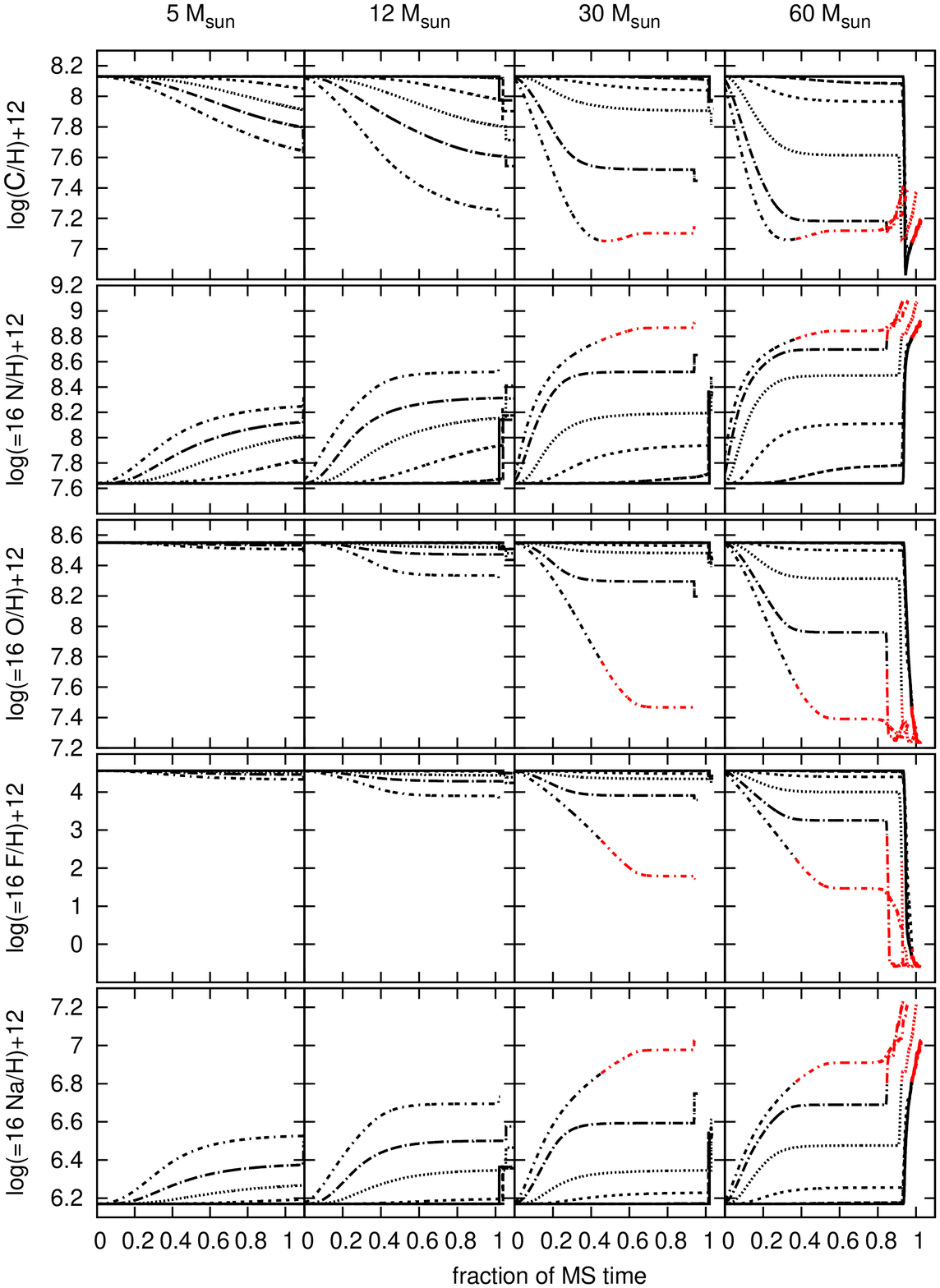}
\caption{continued: Like Fig.~\ref{fig:track_abundance_changes1_smc}, for stellar models of Galactic composition.}
\label{fig:track_abundance_changes2_mw}
\end{figure*}

%-----Surface abundances along isochrones--------------------------------------------------
\begin{figure*}[p]
\centering
\includegraphics[angle=0,width=0.9\textwidth, bb= -10 5 580 800, clip]{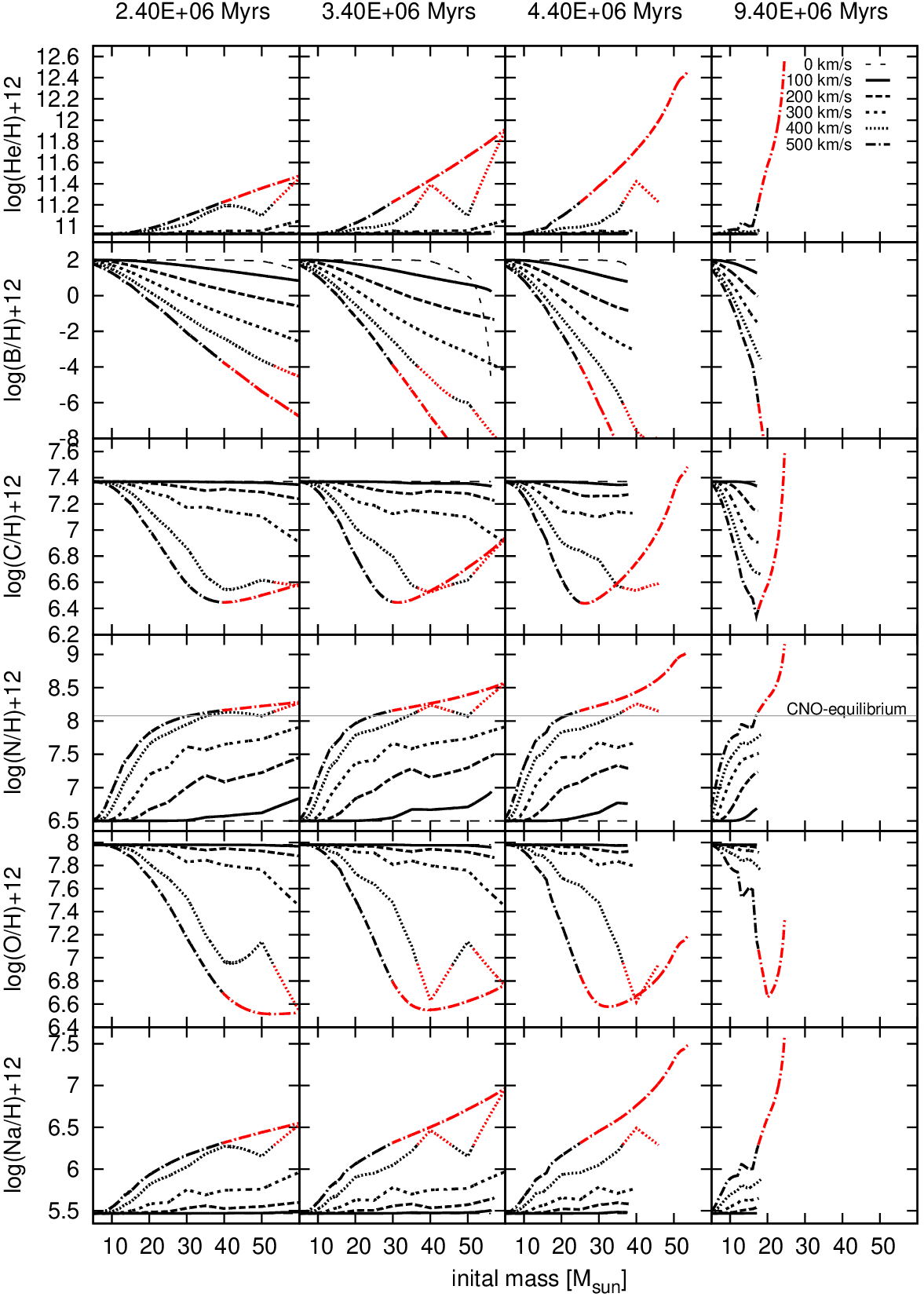}
\caption{Surface abundances for a fixed age, as a function of initial mass and rotational velocity, for SMC metallicity. 
From top to bottom are shown helium, boron, carbon, nitrogen, oxygen and sodium abundances.}
\label{fig:aiso_abundance_changes_smc}
\end{figure*}

\begin{figure*}[p]
\centering
\includegraphics[angle=0,width=0.9\textwidth, bb= -10 5 580 800, clip]{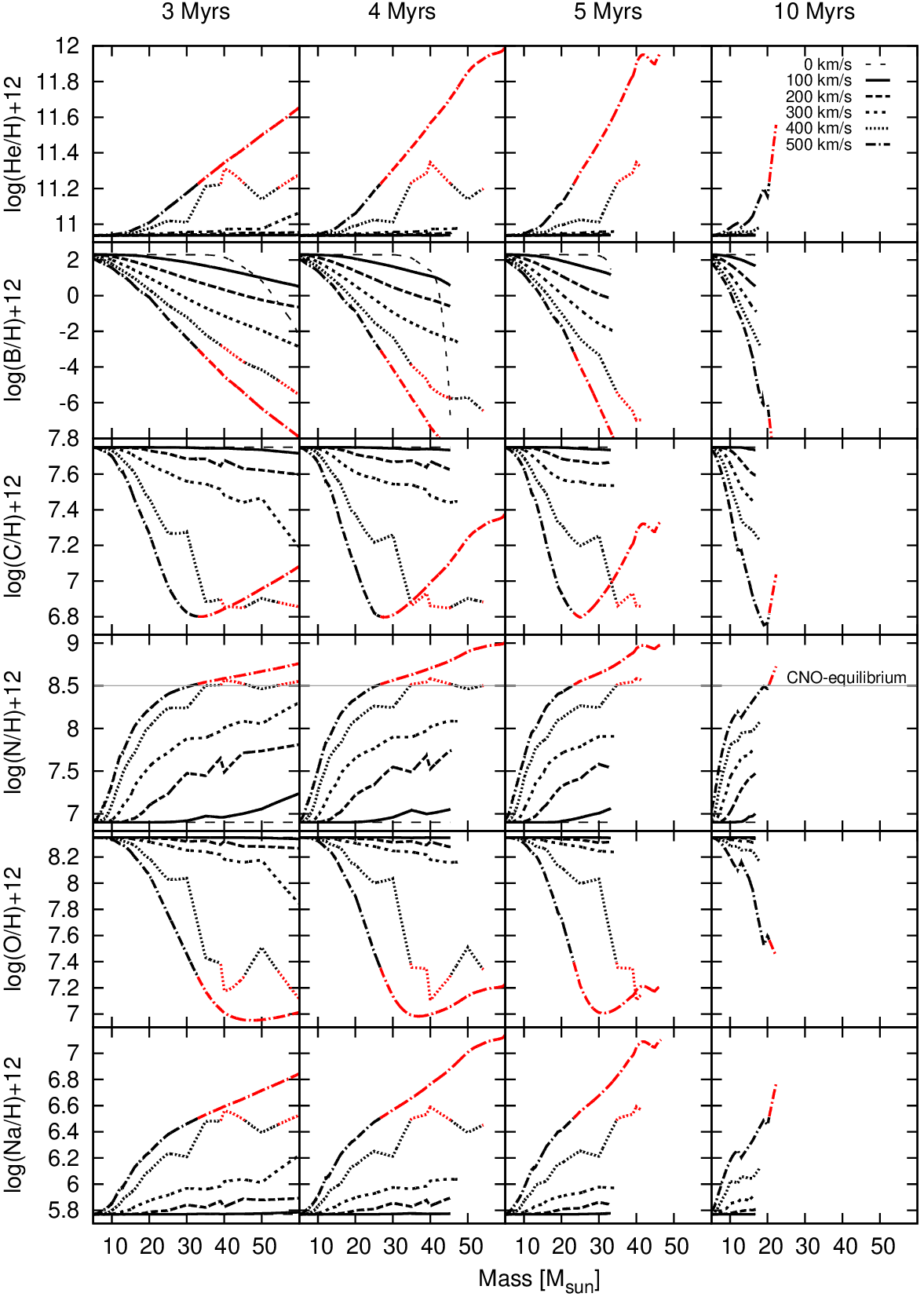}
\caption{Surface abundances for a fixed age, as a function of initial mass and rotational velocity, for LMC metallicity.  
From top to bottom are shown helium, boron, carbon, nitrogen, oxygen and sodium abundances.}
\label{fig:aiso_abundance_changes_lmc}
\end{figure*}

\begin{figure*}[p]
\centering
\includegraphics[angle=0,width=0.9\textwidth, bb= -10 5 580 800, clip]{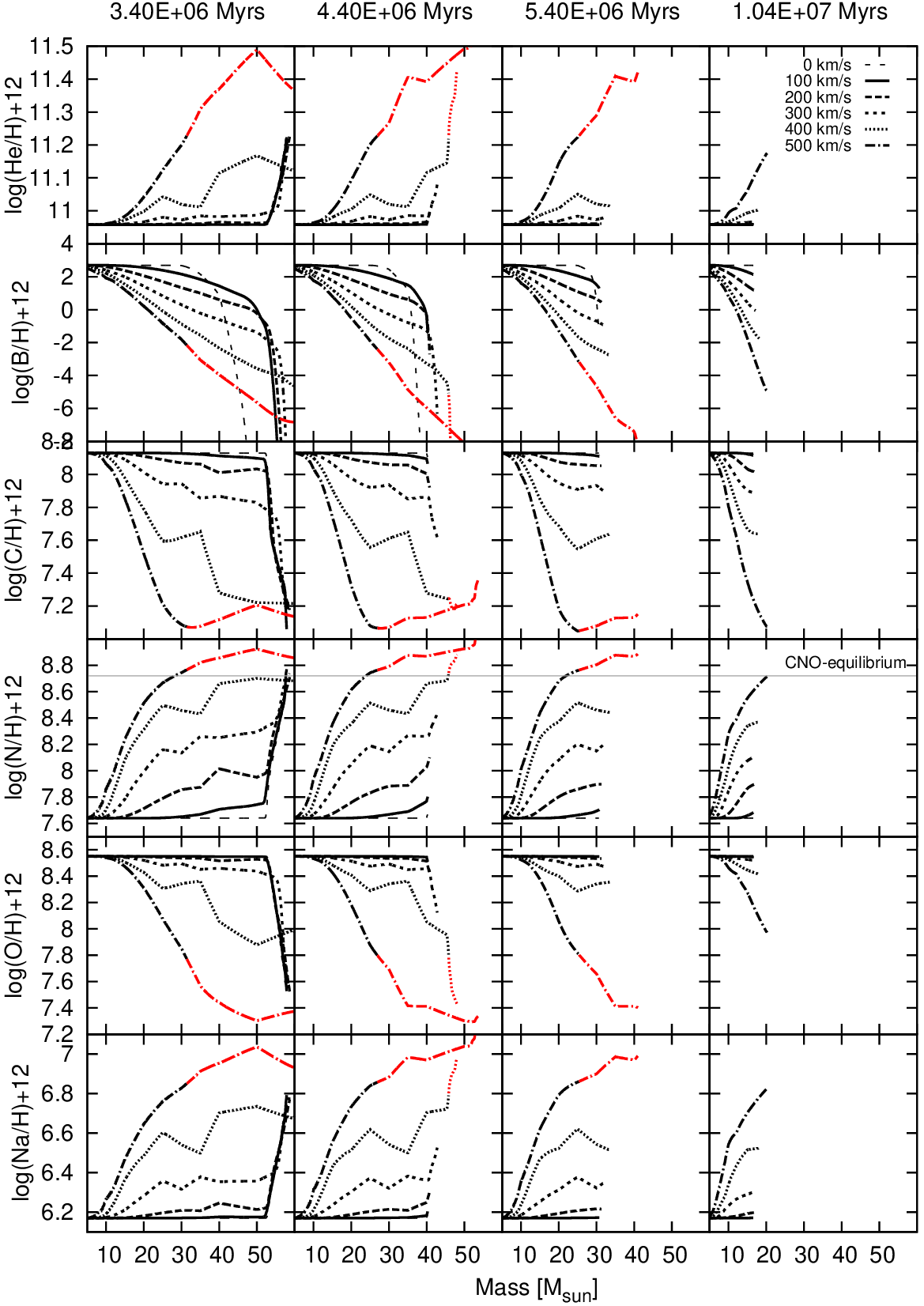}
\caption{Surface abundances for a fixed age, as a function of initial mass and rotational velocity, for Galactic metallicity.
From top to bottom are shown helium, boron, carbon, nitrogen, oxygen and sodium abundances.}
\label{fig:aiso_abundance_changes_mw}
\end{figure*}

\end{document}